\documentclass{aa}
\usepackage{url}
\usepackage{amsmath}
\usepackage[english]{babel}
\usepackage{mathtools}
\usepackage[varg]{txfonts}

\newcommand{\sectionref}[1]{Sect. \ref{#1}}
\newcommand{\figureref}[1]{Fig. \ref{#1}}

\newcommand{\tableref}[1]{Table \ref{#1}}

\newcommand{\cmacionize}{\textsc{CMacIonize 2.0}}
\newcommand{\cmacionizeone}{\textsc{CMacIonize 1.0}}

\begin{document}

\title{\cmacionize{}: a novel task-based approach to \\ Monte Carlo
radiation transfer}

\author{B. Vandenbroucke
        \inst{1,2}
        \and{}
        P. Camps
        \inst{1}}
\institute{Sterrenkundig Observatorium, Universiteit Gent, Krijgslaan 281,
B-9000 Gent, Belgium
           \email{bert.vandenbroucke@ugent.be}
           \and{}
           SUPA, School of Physics \& Astronomy, University of St Andrews,
North Haugh, St Andrews, KY16 9SS, United Kingdom\\}

\date{Received xxx/Accepted xxx}

\abstract
{
Monte Carlo radiative transfer (MCRT) is a widely used technique to model the interaction between radiation and a medium, and plays an important role in astrophysical modelling and when comparing those models with observations.
}
{
In this work, we present a novel approach to MCRT that addresses the challenging memory access patterns of traditional MCRT algorithms, which hinder optimal performance of MCRT simulations on modern hardware with a complex memory architecture.
}
{
We reformulate the MCRT photon packet life cycle as a task-based algorithm, whereby the computation is broken down into small tasks that are executed concurrently. Photon packets are stored in intermediate buffers, and tasks propagate photon packets through small parts of the computational domain, moving them from one buffer to another in the process.
}
{
Using the implementation of the new algorithm in the photoionization MCRT code \cmacionize{}, we show that the decomposition of the MCRT grid into small parts leads to a significant performance gain during the photon packet propagation phase, which constitutes the bulk of an MCRT algorithm, as a result of better usage of memory caches. Our new algorithm is a factor 2 to 4 faster than an equivalent traditional algorithm and shows good strong scaling up to 30 threads. We briefly discuss how our new algorithm could be adjusted or extended to other astrophysical MCRT applications.
}
{
We show that optimising the memory access patterns of a memory-bound algorithm such as MCRT can yield significant performance gains.
}


\maketitle

\section{Introduction}

In the mid 2000s, the clock speed of a single CPU core reached its limits,
imposed by the physics of heat dissipation in increasingly smaller and denser
circuits. Since then, increased CPU performance has been delivered by increased
parallelism rather than a raw increase in speed. Scientific progress through
increased performance started to crucially depend on scalability, a concept that
was relatively unimportant in astrophysical computing until then. The impact
on Monte Carlo radiative transfer simulations can be seen by
comparing historical works. While \citet{1969Price} mentions a reference speed
of $10^6$ photon packets in 200~minutes for an 18 cell grid,
\citet{2004Wood} quote a speed of $10^6$ photon packets per minute for a
similar algorithm on a $65^3$ grid, or a speedup of roughly $3\times{}10^6$.
\cmacionizeone{}, a rewrite of the \citet{2004Wood} code manages $10^7$
photon packets per minute for a $64^3$ grid on a single CPU core, or a speed up
of less than 10, well below the expected factor of $\approx{}200$ if Moore's law
for single CPUs would still hold. This factor can be recovered by running
\cmacionizeone{} on $\approx{}30$ cores, a representative number for a
modern computing node.

While the speed of CPUs increased and ultimately hit a limit, memory went from
being a limiting factor (\citealp{1969Price} mentions that a 20 cell grid was
impractical, presumably for this reason) over being abundant to being a limiting
factor again. Modern computers support memory sizes of
$10-100~{\rm{}GB}$ and with that grid sizes of up to $1000^3$ cells (every
double-precision floating point number
stored in a cell of a $1024^3$ cell grid corresponds to 8~GB of memory space
required). This memory, however, is generally laid out in a complex way, with most of
the memory having a relatively low bandwidth. This results in algorithms that
are severely memory-bound, i.e. the speed of the algorithm is dominated by the
memory bandwidth rather than the clock speed of the CPU cores. The speed of
these algorithms crucially depends on memory access patterns that can lead to
efficient cache usage, concepts that are still relatively new in astrophysical computing.

These two factors (increased parallelism and increased memory boundedness) have
made it increasingly hard to develop astrophysical software in the last
decade. Successful algorithms have required a radical redesign,
often depending on an active
involvement of computer scientists in the development process
\citep{2012Bordner,2013Gonnet, 2016Schaller, 2016White, 2018Nordlund,
2018Borrow}. Given the current trends in hardware technology, 
this way of developing astrophysical software may become the norm.

In this work, we present a novel approach to the algorithmic framework of the
Monte Carlo radiative
transfer (MCRT) technique, which is used for radiation transport simulations
in various fields, including, for example, photoionization
\citep{2016Hubber,2018Vandenbroucke_SILCC},
dust continuum radiation \citep{2013Steinacker,2020_SKIRT9},
resonant line transfer \citep{2015_Smith,2018_Behrens}
and particle physics \citep{2003_MCNP,2003_Geant4}.
We reformulate the photon packet
propagation mechanism at the heart of the MCRT technique as
a task-based algorithm. The algorithm is defined in
terms of individual units of computational work, each involving small
amounts of memory and with clear inter-dependencies 
governing which tasks can be executed at what time and which tasks can be executed
concurrently.

The major bottleneck in contemporary MCRT algorithms is the grid
structure used to discretize the interstellar medium in space. In multi-node, distributed
memory setups, it is easily understood that 
this grid structure leads to significant overheads, either
because the entire  structure needs to be duplicated on all nodes
\citep{2018Vandenbroucke_CMacIonize,2020_SKIRT9}, or because photon packets need to be
communicated between nodes \citep{2019Harries}. On shared memory systems, the
grid is, however, a bottleneck as well, this time due to bandwidth issues. The grid occupies a
large (possibly continuous) chunk of memory that is accessed concurrently
by different threads. Due to the random nature of the MCRT algorithm,
these concurrent access events happen in an arbitrary fashion that puts
significant strains on the memory bandwidth, suffers from generally
unpredictable retrieval delays and makes it nearly impossible for algorithm
implementations to efficiently use fast memory caches.

The obvious solution to these issues is to give up on the idea of a single
monolithic grid structure, even in the case of a shared memory algorithm, and
instead decompose the grid into many smaller parts, which we will call
\emph{subgrids}. These subgrids can then act as the small memory resources on
which the tasks in the task-based algorithm operate. As in the case of a
distributed-memory grid, this inevitably means storing photon packets that
transition from one such subgrid to a neighbouring subgrid, which requires the
use of \emph{buffers} \citep{2019Harries}.

\begin{figure}
\centering{}
\includegraphics[width=0.48\textwidth{}]{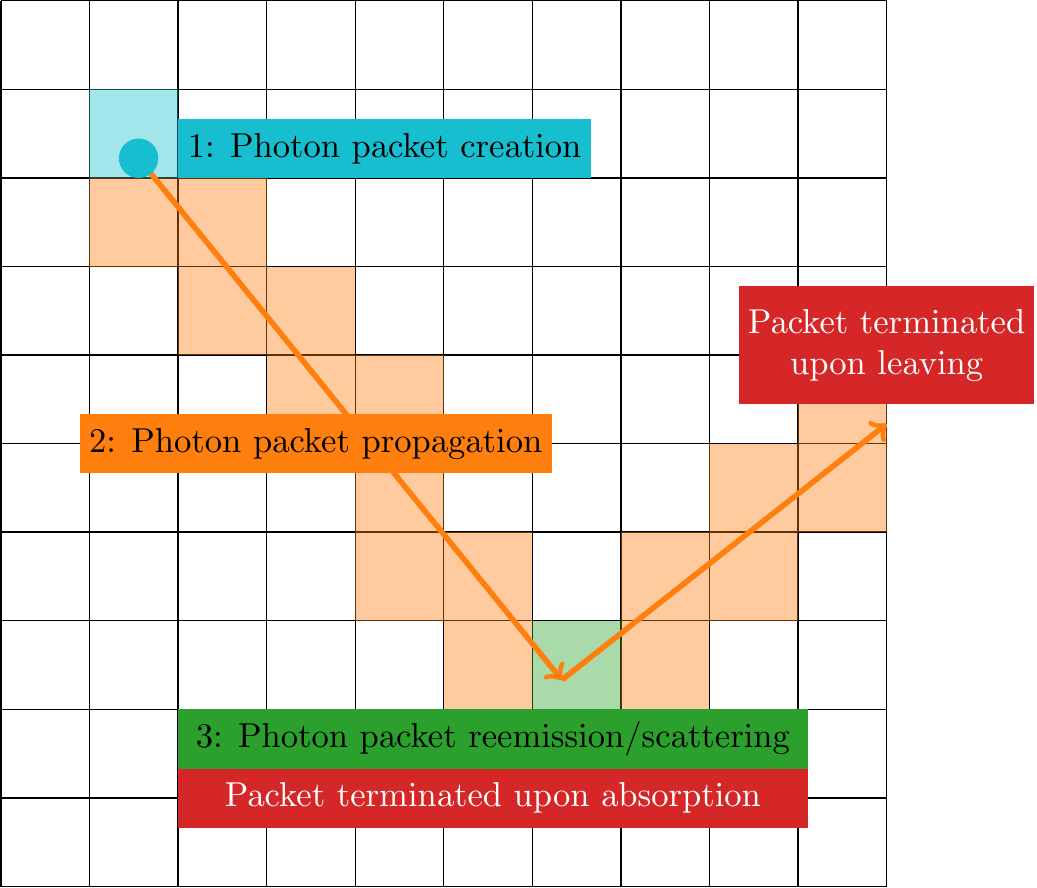}
\caption{The basic life cycle of a Monte Carlo photon
packet. The photon packet is generated by a source and emitted in a randomly
sampled direction. It is then propagated through the grid structure until a
predetermined randomly sampled optical depth has been reached. Upon absorption,
a randomly sampled fraction of photon packets can undergo re-emission or
scattering, which will alter the properties of the photon packet. A photon
packet's life cycle is terminated when the photon packet is absorbed and not
re-emitted or scattered, or if it leaves the boundaries of the system.}
\label{fig:mc_overview}
\end{figure}

Any minimal task-based MCRT algorithm hence depends on
three crucial ingredients: an efficient strategy to decompose the computational
grid into independent subgrids, the formulation of a set of tasks that
perform photon packet propagation based on interactions
between these subgrids and the corresponding buffers, and
an algorithm to create and schedule these tasks.
At the same time, the task-based approach does not
require any changes to the implementation of the physical
equations governing the simulation. The interaction between the photon packets
and the medium represented by each of the cells in the individual subgrids 
can still be computed in the same way as for a single grid structure.
What needs to change is the idea
of a monolithic photon packet life cycle, whereby a photon packet is
tracked sequentially from its generation at a source location, throughout its
propagation through the grid structure, until it is terminated by an absorption
event or because it leaves the computational domain. An example of this basic
life cycle is depicted in \figureref{fig:mc_overview}.
In the new task-based algorithm, photon packet
generation and propagation, as well as optional additional processes like
scattering and re-emission, are treated separately and not necessarily
consecutively.

We introduce the task-based algorithm in \sectionref{sec:algorithm} using
the specific example of photoionization as it is implemented in \cmacionize{}\footnote{The source code of \cmacionize{} is hosted on \url{https://github.com/bwvdnbro/CMacIonize}},
the new version of the code presented in \citet{2018Vandenbroucke_CMacIonize}.
This allows us to focus on the basics of the task-based algorithm, without the
need to expand into extra complexities. For the same reason, we assume a regular grid
structure that is subdivided into regular subgrids with identical dimensions, and
we limit the discussion to one or more point sources. In
\sectionref{sec:validation} we validate this task-based algorithm, and in
\sectionref{sec:performance} we compare its performance with that of a more
conventional MCRT photoionization algorithm, i.e.\ \cmacionizeone{}. We also discuss practical aspects
such as the memory requirements of the algorithm and its various components.
Possible extensions of the algorithm are discussed in
\sectionref{sec:extensions}. We highlight some features of
\cmacionize{} that go beyond the basic task-based algorithm, and discuss
how the task-based algorithm could be extended for other MCRT applications.
In \sectionref{sec:conclusion}, we
end with a summary and some overall conclusions.

\section{Task-based algorithm}
\label{sec:algorithm}

\begin{figure*}
\centering{}
\includegraphics[width=0.98\textwidth{}]{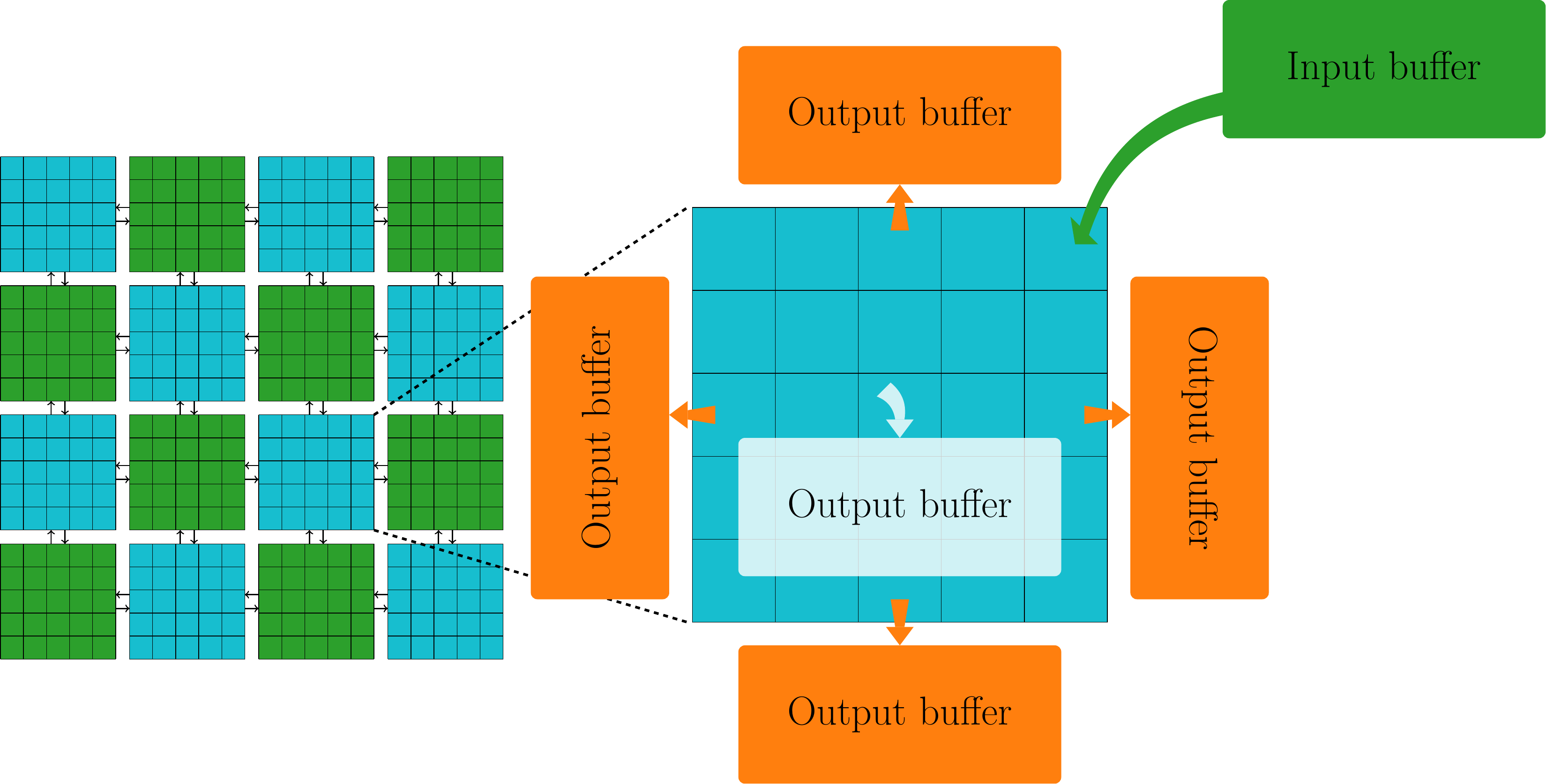}
\caption{The distributed grid used by the task-based
algorithm. Each subgrid is treated as an independent data
structure. The subgrids are only aware of the larger structure of the grid
through information about their neighbours, as indicated by the
bidirectional arrows linking the different subgrids. Photon packets enter the
subgrid through an input buffer and are stored in output buffers according to
their interaction with the subgrid.}
\label{fig:subgrids}
\end{figure*}

\subsection{Calculating photoionization}

The aim of the photoionization algorithm is to determine the
equilibrium temperature and ionization state of an interstellar medium that is
exposed to a radiation field with a known ionizing spectrum. Depending
on the chemical elements that are included and the cooling and heating processes that are
considered, this problem can be arbitrarily complex. We here describe the
example of a hydrogen-only gas, which captures all the essential aspects of the
algorithm.

Within each cell of the computational domain, the composition of the hydrogen
plasma is characterised by two variables: the hydrogen number density, which is
assumed to remain constant during the simulation, and the neutral hydrogen  fraction,
which encodes
the ionization state of the medium within the cell. The latter is computed
by balancing the number of ionization events per unit time, caused by the absorption of
ionizing radiation, with the collisional recombination rate,
which depends on the electron temperature and the free electron density. The
temperature, in turn, can be computed by balancing the photoionization
heating rate with an appropriate cooling rate. If the characteristics of the
embedding photoionizing radiation field are known, one can self-consistently solve
for both the neutral fraction (and electron density) and the temperature.

To model the interaction between the radiation field and the medium in each cell,
the radiation is discretized using a (large) number of \emph{photon packets}. Each
photon packet carries a fixed photoionization rate, and is emitted in a randomly
sampled direction and with a randomly sampled frequency, which determines its
photoionization cross section. Both random values are sampled from appropriate
distribution functions; emission is usually assumed isotropic, while frequencies
are distributed according to a source spectrum.

The photon packets are propagated through the grid until a randomly sampled
optical depth is reached (distributed exponentially). The optical depth along
the path is computed from the path lengths traversed by the photon packet in
each cell, using the neutral hydrogen density in each cell and the
photoionization cross section of the photon packet. These same path lengths and
cross sections are also accumulated in each cell for each passing photon packet,
providing an estimate of the absorbed ionizing radiation field.

When a photon packet reaches its target optical depth, it is absorbed, i.e.\ it
is assumed that all physical photons in the packet have been absorbed along the
way. At this point, the photon packet is reused to sample the diffuse radiation
field caused by ionizing recombinations. For a solar-metallicity \textsc{Hii} region with
a temperature of $8,000$~K, $36~\%$ of the photons emitted by recombination
events have frequencies above the ionization threshold of hydrogen and 
will contribute to the ionizing radiation field. A corresponding fraction
of the absorbed photon packets is therefore re-emitted in a randomly
sampled new direction and with a randomly sampled new frequency. Since in our
algorithm this re-emission happens instantaneously, it is essentially treated as a
scattering event.

At the end of the photon packet propagation phase, each cell holds an
estimate of the absorbed ionizing radiation field,
which is then used to compute the temperature and neutral fraction in the cell.
Since the neutral fraction determines the neutral hydrogen density, which in
turn sets the opacity of the cell for ionizing radiation, the
whole process must be repeated until a converged solution is obtained. Starting from a
highly ionized medium that is transparent for ionizing radiation, 
convergence is typically reached within 10 iterations
\citep{2018Vandenbroucke_CMacIonize}. Convergence is much slower when starting
from an assumed neutral initial medium.

\subsection{Tasks}
\label{sec:tasks}

Within our task-based photoionization algorithm, the conventional photon packet
life cycle is broken up as follows (see \figureref{fig:mc_overview}):
\begin{itemize}
\item{} Photon packet generation: a number of photon packets is generated
sequentially from the source model and placed in an input buffer for the subgrid
containing the source location.
This buffer is then flagged for processing, while the
thread that executed this work moves on to do something else.
\item{} Photon packet propagation: the photon packets in a subgrid's input
buffer are propagated
through the subgrid, and are placed in buffers according to their interaction
with that subgrid (see \figureref{fig:subgrids}).
Photon packets that leave the subgrid through one of its
boundaries are passed on to buffers for neighbouring subgrids, while photon
packets that are absorbed internally are tracked in another buffer. Photon
packets that leave the subgrid through a boundary that is also a simulation box
boundary are terminated.  During
this step, the path length counters in the cells that keep track of the ionizing
radiation field are updated.
\item{} Photon packet re-emission: photon packets that were put in a subgrid
buffer after absorption are re-emitted according to the local properties of the relevant
subgrid cell. In practice, this means that the properties (energy, propagation
direction) of the photon packets are updated inside the buffer. Photon packets
that are not eligible for re-emission are terminated and removed from the buffer.
\end{itemize}

Any of these steps can be executed concurrently by
different threads, as long as causality is respected for individual photon
packets. Each step represents a \emph{task}: a small amount of
computational work with a clearly defined memory footprint and known
dependencies. The efficient scheduling and execution of these tasks is at the
core of the task-based algorithm. All tasks are typically put into one or more
queues. Parallel execution threads query those queues for available tasks
and execute them, until all tasks have been executed. Tasks are added to the
queues by a \emph{scheduler} that makes sure causality for tasks is respected
\citep{2016Gonnet}. By using a separate queue for each thread, it is possible
to sort tasks so that threads
preferentially execute tasks accessing the same data structures to help increase cache
efficiency.

In conventional task-based algorithms, the tasks and their inter-dependencies are
known at the start of the calculation \citep{2016Gonnet}. In our
task-based MCRT algorithm, however, the set of tasks is not a priori known because of
the random nature of the algorithm. When a task finishes, it can
\emph{create} a new task. This leads to the following \emph{implicit} task dependencies:

\begin{itemize}
\item{} An initial setup loop creates photon packet generation tasks for all 
sources in the model.
\item{} A photon packet generation task creates a single photon packet
propagation task that acts on the buffer that was filled by the 
generation task and on the subgrid that contains the position of the point
source that emitted the photons.
\item{} A photon packet propagation task creates new 
propagation tasks and up to one re-emission task when the respective
buffers containing the outgoing photon packets are full. These new tasks
act on the corresponding output buffers and the subgrid associated with it: the
corresponding neighbouring subgrid for neighbour buffers, and the original
subgrid for the internal output buffer.
\item{} A photon packet re-emission task similarly creates up to one 
propagation task.
\end{itemize}

\subsection{Subgrids and photon packet buffers}
\label{sec:subgrids_buffers}

The spatial grid in our algorithm is stored as an array of pointers to
individual subgrid structures that are completely self-contained. Each subgrid stores
 the geometry of its cells plus all of
 the physical quantities and variables for these cells. For a regular
grid, the cell geometry can be represented by just the anchor point and the side
lengths of the cuboid that encompasses the subgrid. 
Furthermore, a subgrid also stores information about
its position within the larger grid, i.e.\ it stores the indices
of the neighbouring subgrids in the array. This is illustrated with arrows in
the left panel of \figureref{fig:subgrids}.

Most of the MCRT algorithm is implemented as functions that directly
manipulate an individual subgrid. The photon propagation step takes an input
buffer, propagates all the photon packets it contains through the subgrid,
and stores the resulting outgoing photon packets in appropriate output buffers.
The re-emission step uses local cell variables to decide if and how to
re-emit a photon packet that was locally absorbed earlier. These parts of the
algorithm are almost identical to the corresponding parts in a non task-based
version of the algorithm.

Each buffer contains a fixed number of photon packet slots and a counter
indicating the actual number of photon packets currently stored in the buffer.
It also stores
general directional information about the photon packets in the buffer that is
used to speed up the continued propagation in other subgrids. For example, if the buffer
collects photon packets that leave through the front face of the subgrid cuboid,
then these photon packets will all enter the neighbouring subgrid through the
back face.

Each photon packet stores all the information required to propagate the packet
through the grid. This includes the propagation direction, the current position, the optical depth accumulated so far, the target optical
depth, and the energy/frequency and/or interaction cross sections
required in the optical depth calculation. The idea is that
the propagation of a photon packet can be \emph{paused} at any time during its life cycle.

The task scheduler provides a pool of pre-allocated photon packet buffers
that are assigned and reassigned as needed during the operation of the algorithm.
Specifically, each subgrid is assigned an output buffer corresponding to each of
its neighbouring subgrids in addition to a single re-emission buffer. Furthermore,
every source is assigned an output buffer as well. In each case, whenever 
a task fills up or otherwise completes an output buffer, the buffer is passed 
on to the scheduler to serve as an input buffer for another task, and 
a new empty buffer is acquired to replace the original output buffer.

\subsection{Queues and scheduling}
\label{sec:queues}

The task scheduler manages a number of queues to track tasks that are ready to be executed. 
The information required to execute a particular task is stored in a small data structure
including the type of the task and the indices of the resources involved
in its execution. For example, a photon propagation task references the 
relevant subgrid and input photon buffer.

\begin{figure}
    \centering{}
    \includegraphics[width=0.48\textwidth{}]{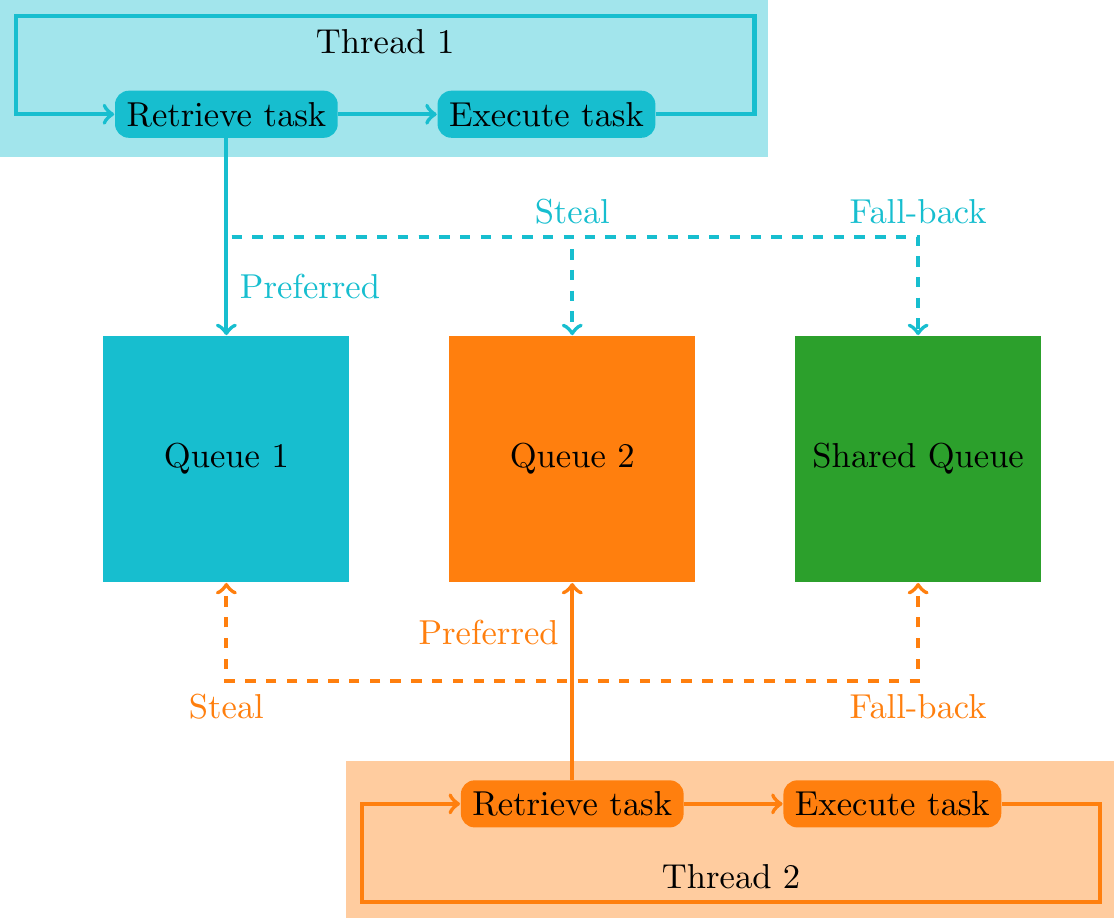}
    \caption{Task retrieval for a simulation with two parallel execution threads. Both threads retrieve tasks from their preferential queue unless this is not possible. In this case, they first try to steal a task from the other thread's queue, and if this does not work, they retrieve a task from the shared queue.}
    \label{fig:queue_mechanism}
\end{figure}

Our algorithm uses several queues. 
Each execution thread is assigned a \emph{per-thread} queue  that
stores the propagation tasks for the
subgrids that are often accessed by that specific thread and from which that
thread preferentially gets its tasks. Moreover, there is a single
\emph{shared} queue that stores the
photon packet generation and re-emission tasks that are only
executed if no propagation tasks are available.

As illustrated in \figureref{fig:queue_mechanism}, during the task execution loop, a thread
\begin{enumerate}
\item{} queries its own queue for a propagation task and executes it, or, if
this fails,
\item{} tries to steal a propagation task from another thread's queue and
executes it, or, if this fails,
\item{} tries to acquire a re-emission or photon generation task from the shared
queue and executes it, or, if this fails,
\item{} tries to prematurely schedule a non-full outgoing photon buffer for one
of the subgrids.
\end{enumerate}

The order of these actions has been carefully chosen to maximise the
throughput of photon packets, since this minimises the number of
buffers required to store photon packets that have not yet been terminated.
The last action in the above list is necessary to guarantee that the
algorithm will finish. Indeed, output buffers
associated with individual subgrids will not necessarily be full by the time all photon
packets have been injected into the grid, so that an alternative
method is required to ensure that packets in partially filled buffers are
properly processed.

Because the shared queue contains all photon packet generation tasks, buffers
can only be prematurely scheduled once all photon packets have been generated.
Even then, care must be taken not to reduce the efficiency of the algorithm by
prematurely scheduling too many partially filled photon packet buffers when 
the photon packet propagation phase is winding down. This is achieved by
preferentially scheduling buffers with a larger number of photon packets
as long as they are available.

If a thread cannot obtain an executable task after performing all of the actions in the
above list, then there is a possibility that all photon packets have been
successfully propagated through the  grid, and an appropriate check
is performed to detect this. If this check fails, the thread repeats the
scheme above in case  a new or existing task has become
eligible for execution. The loop continues until the thread detects a
successful termination of the propagation phase. At this point all threads
will agree that
the propagation phase finished and will exit the parallel environment.

After the propagation phase finishes, additional subgrid tasks can be executed,
such as the calculation of the ionization balance in each cell. Generally, these
tasks can again be performed in parallel. When appropriate, the photon
propagation and cell update phases are repeated until convergence has been reached.
The exit condition for each phase inevitably introduces a synchronisation point within the parallel execution where load imbalances lead to idle time for some of the threads. 
However, as long as the number of phase iterations is
small compared to the number of tasks performed within each phase, 
the total load imbalance should be small.

\subsection{Subgrid copies}
\label{subgrid_copies}

An important feature of our algorithm is the ability to duplicate
subgrids for which there is a high contention. In many simulated models,
some subgrids have a significantly higher-than-average computational load.
A good example is a model with a single point source of radiation. In such a
simulation, the subgrid containing the source position will receive all of the
initial photon packet propagation tasks. 
Since each photon packet propagation task needs unique access to this
subgrid, this leads to a very strong bottleneck early in the simulation.

\begin{figure}
\centering{}
\includegraphics[width=0.45\textwidth{}]{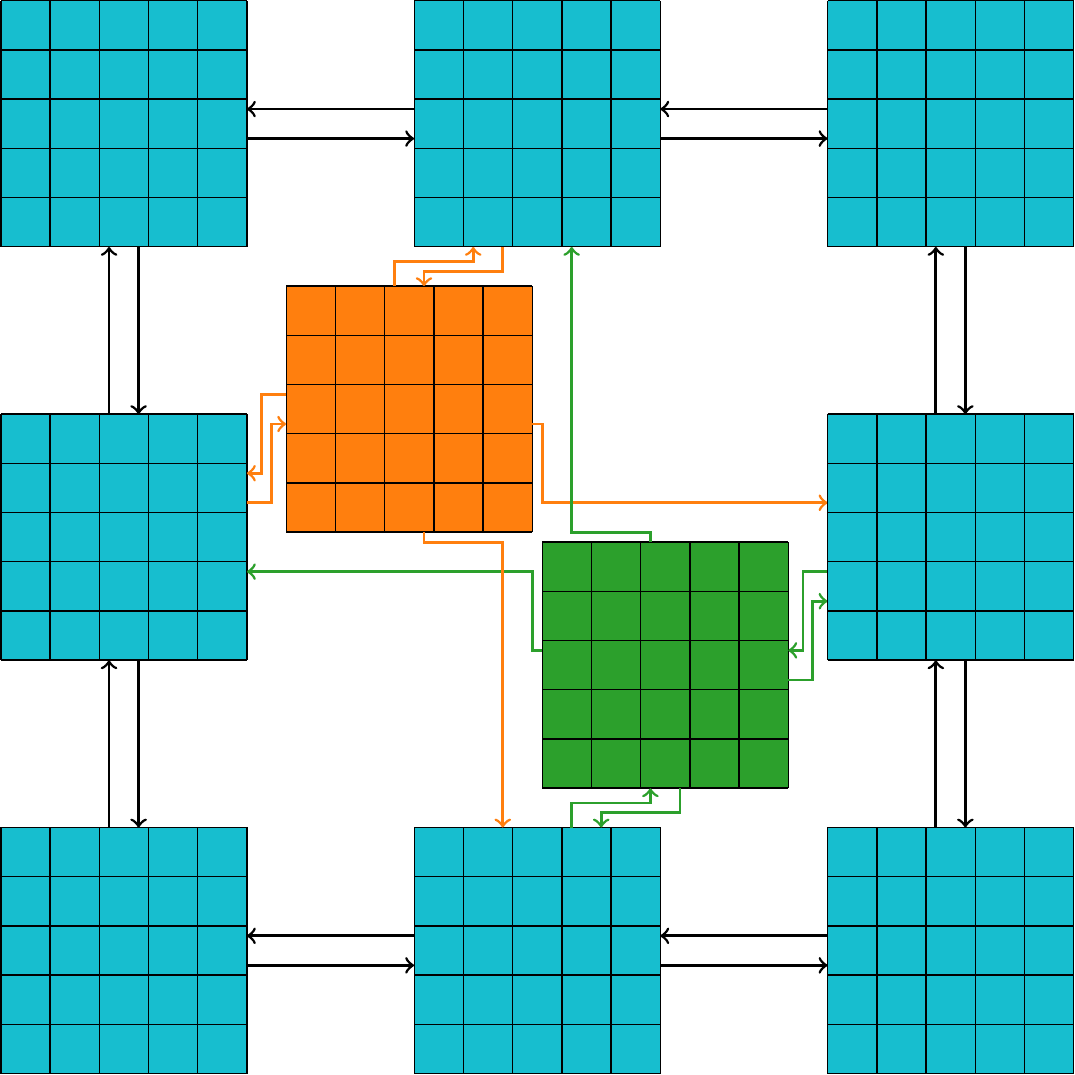}
\caption{A distributed grid with subgrid copies. The subgrids in
green and orange represent the same portion of the grid, but are in no way
linked during the photon propagation step. They each store information about
their neighbours as if they were a regular subgrid. However, their neighbours only
store information about one of the copies, so that the neighbours themselves are
also unaware of the subgrid copy. Apart from actually creating the subgrid copy,
this rerouting of the subgrid interconnections is the only step that is required
to use the subgrid copies in the task-based algorithm and imposes very little
overhead.}
\label{fig:subgrid_copies}
\end{figure}

This problem can be addressed by making one or more copies of the 
subgrids for which a lot of contention is expected. These copies are
added to the list of subgrids as independent subgrids, but inherit the positions
and optical properties of their parent subgrid. The neighbouring relations
between the original subgrid, its neighbours, and the copies are
rearranged so that each subgrid still has an outgoing neighbour on all sides (see
\figureref{fig:subgrid_copies}). Specifically, each of the neighbouring subgrids stores
an outgoing link to just one of the subgrid copies, distributed evenly over the copies to help
balance their workload. 
Each subgrid copy keeps track of its own radiation field
counters and behaves as if it was not related to its parent.
At the end of the photon propagation phase, the counters for all copies are
accumulated into the parent counters and all copies are discarded.

If the individual subgrids are sufficiently small, the copy procedure incurs little
overhead, while significantly improving the load balancing of the algorithm.
However, a successful duplication strategy requires
a good prediction of the computational load of each subgrid.
One might obtain such an estimate based on a previous simulation run
or iteration or based on a low resolution bootstrap simulation. 
While these options could in principle provide an accurate prediction, they are
often impractical. We found that a heuristic approach is easier to use and
provides a sufficiently accurate estimate for our purposes.

\begin{figure}
\centering{}
\includegraphics[width=0.48\textwidth{}]{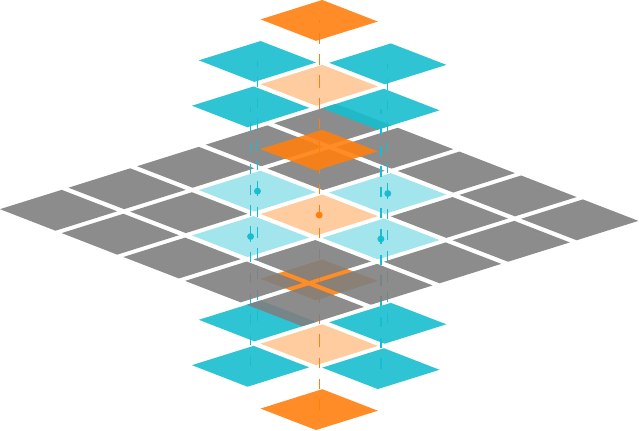}
\caption{The copy hierarchy used to balance the computational load
for subgrids with a large contention. The subgrid depicted in orange contains a
source, has a copy level of 2 and is surrounded by 4 subgrids with a copy
level of 1 that are in turn surrounded by normal subgrids with a copy level of
0. In this example, 7 additional subgrids are created that each act as
independent subgrids during the photon propagation step.}
\label{fig:copies}
\end{figure}

The heuristic we chose to adopt assumes that there is a strong correlation
between the presence of a source within a subgrid and an increased load within
that subgrid and its neighbours. Each subgrid is assigned a \emph{copy level},
where a copy level equal to $l$ gives rise to $2^l$ duplicates (including the
original). The user specifies a single parameter, $l_\mathrm{max}$, indicating
the copy level for subgrids that contain a
source position. Once this maximum copy level has been assigned to all subgrids associated
with a source, the heuristic recursively
traverses the subgrid structure and ensures that neighbouring subgrids are at
most one copy level apart. If a given subgrid has level $l_i$, then all of
its neighbours have levels $l_i - 1 \leq{} l_j \leq{} l_i + 1$. This
automatically leads to a hierarchical structure of duplicates that captures the
expected computational load for a uniform density distribution close to a
source (see \figureref{fig:copies}), and we found that this approach works well
in all our tests.

\subsection{Memory management and locking}
\label{sec:memory_management}

The multiple execution threads of our algorithm operate in a single,
shared memory environment. It is therefore necessary to provide the
appropriate synchronisation mechanisms for accessing common data structures.
A key benefit of the subgrid approach presented in the previous sections
is that most of the data  are local to the corresponding subgrid. 
As long as at most a single task operates on a given subgrid at any one time,
there is no need to synchronise access to these local data.
The data structures that do need synchronised access thus include the 
subgrids, the photon packet buffers (see \sectionref{sec:subgrids_buffers}),
the task definitions, and the task queues (see \sectionref{sec:queues}).
All synchronisation occurs at the level of the task scheduler and never
within any of the tasks themselves.

To begin with, each of the queues managed by the task scheduler is protected
by a single lock that must be acquired by a thread before it can retrieve a new task
from the queue. As described in \sectionref{sec:queues}, each thread has its own
preferential queue and only occasionally accesses the other queues.
As a result, contention on the queue locks is low and the overhead is minimal.

The situation is more complicated for the data structures holding the information
describing each task and for the photon packet buffers. Our algorithm continuously
creates new tasks and terminates other tasks as the simulation proceeds (see \sectionref{sec:tasks}). Similarly, photon packet buffers are continuously being
assigned and reassigned (see \sectionref{sec:subgrids_buffers}).
To minimise the overhead of this dynamic process, an array of empty task descriptions
and an array of empty photon packet buffers are pre-allocated before parallel task
execution begins. The main drawback of this approach is that the arrays have
a fixed size and thus must be chosen sufficiently large to support the requirements
of the simulation. It is therefore essential to have a good heuristic for predicting
these requirements. We will return to this issue in \sectionref{sec:performance}.

Because task descriptions and photon packet buffers are frequently accessed
from all execution threads, protecting access with a single, global lock for 
each array would cause substantial overhead. Instead, each entry in these arrays
is equipped with its own lock to allow fine-grained access control. In addition,
a single \emph{atomic} index counter (per array) is used to help locating available 
entries. Specifically, whenever a thread needs to acquire a new entry, it
atomically increments the index counter, computes an actual array index from the
counter by taking its value \emph{modulo} the array size, and then attempts
to acquire the lock for the corresponding entry. If this is unsuccessful,
the entry is already in use and the whole procedure is repeated.
Because entries can be expected to be released in roughly the same order as
in which they were acquired, this rolling index mechanism will often be successful
after one or just a few attempts.

Lastly, all subgrids in the distributed grid structure, including any duplicates
(see \sectionref{subgrid_copies}), are also pre-allocated before parallel task
execution begins, and each subgrid is equipped with a lock that controls
access to it.

As mentioned in \sectionref{sec:queues}, a task description lists the resources
needed by the task. For example, a photon propagation task references the
relevant subgrid and input photon buffer. Following the technique presented by
\citet{2016Gonnet}, a thread that wants to acquire a task tries to acquire a
lock for all the resources listed by the task. If this does not succeed,
any locks that were successfully acquired are released again, and the thread
moves on to look for another eligible task. This mechanism ensures exclusive
access to a task's resources during its operation.

\section{Validation}
\label{sec:validation}

We have validated the new photoionization algorithm using a suite of tests and benchmark runs. The parameter- and initial condition files for these tests are part of the public \cmacionize{} repository. Reference results were already presented in \citet{2018Vandenbroucke_CMacIonize} and have not changed, since no changes were made to the physical equations we are solving. Below, we give a brief overview of these tests, the aspects of the code they test, and any changes made since the release of \cmacionizeone{}.

\subsection{Str\"{o}mgren sphere}

When a single source with an ionising UV luminosity $Q_{\rm{}H}$ photoionizes a uniform, hydrogen-only medium with number density $n_{\rm{}H}$ and a fixed collisional recombination rate $\alpha{}_B$, it will ionize out a sphere with a radius that can be approximated as
\begin{equation}
R_S = \left( \frac{3Q}{4\pi{}n^2_{\rm{}H}\alpha{}_B} \right)^{\frac{1}{3}}.
\label{eq:stromgren_volume}
\end{equation}
If the UV source furthermore emits radiation at a fixed frequency, so that the photoionization cross section $\sigma{}_{\rm{}H}$ can be assumed to be constant, then the ionization balance equation at radius $r$ can be written as
\begin{multline}
Q\sigma{}_{\rm{}H}n_{\rm{}H}\exp{} \left(- n_{\rm{}H} \int_0^r \sigma{}_{\rm{}H}
x_{\rm{}H}(r') {\rm{}d}r'\right) =\\ 4\pi{}r^2n^2_{\rm{}H}\alpha{}_B\left(1 -
x_{\rm{}H}(r)\right)^2,
\label{eq:stromgren_profile}
\end{multline}
with $x_{\rm{}H}(r)$ the hydrogen neutral fraction at radius $r$.

This equation can be numerically solved on a grid in $r$ to provide a semi-analytic reference solution for a photoionization simulation using the same setup. In addition to testing that the correct volume is ionized out (this diagnostic only depends on the recombination rate), this also checks that the width of the ionization front and the exponential extinction of the ionizing radiation is resolved correctly (these diagnostics also depend on the photoionization cross section).

The Str\"{o}mgren volume test is an ideal test case for any basic photoionization algorithm, since it tests all important aspects of the algorithm: the interaction between individual photon packets and the medium, the update of the cell properties at the end of each iteration, and the iterative scheme to obtain a converged solution. More specifically for the new MCRT algorithm, this test checks that all photon packets are correctly passed on from one subgrid to another, that the contributions from subgrid copies are correctly taken into account, and that subgrid copies are correctly synchronised.

The benchmark test contained in the repository uses a $10\times{}10\times{}10~$pc box with a uniform hydrogen-only medium with a density $n_{\rm{}H}=100~{\rm{}cm}^{-3}$. At the centre of the box is a single photoionizing source with an ionizing luminosity $Q_{\rm{}H}=4.26\times{}10^{49}~{\rm{}s}^{-1}$ that emits monochromatic radiation at the hydrogen photoionization energy, $13.6~$eV. We assume a constant photoionization cross section $\sigma{}_{\rm{}H}=6.3\times{}10^{-18}~{\rm{}cm}^2$ and a constant recombination rate $\alpha{}_{\rm{}H}=4\times{}10^{-13}~{\rm{}cm}^3~{\rm{}s}^{-1}$, and consider a version without and with diffuse re-emission. For the latter case, we assume that photons that get absorbed have a constant probability, $P_r=0.36$, to be re-emitted as photoionizing photons. In this version of the test, the semi-analytic reference solution is no longer accurate, but the ionized volume still satisfies equation \eqref{eq:stromgren_volume}, albeit with a different value for the ionizing luminosity, $Q'=Q/(1-P_r)$.

\begin{figure}
    \centering
    \includegraphics[width=0.48\textwidth{}]{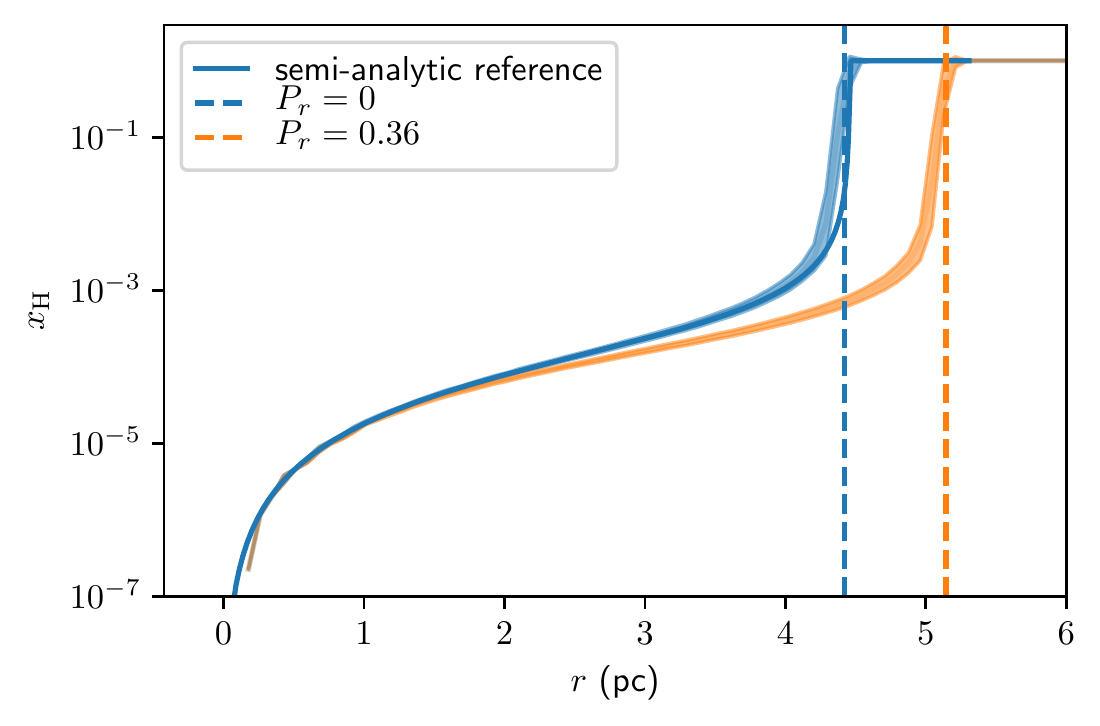}
    \caption{Hydrogen neutral fraction $x_{\rm{}H}$ as a function of radius $r$ for the Str\"{o}mgren volume tests. The simulation results were binned in 100 radial bins in log space. The full lines show the average in each bin, while the shaded regions underneath the curves correspond to the standard deviation in each bin. For reference, the semi-analytic reference solution derived from equation \eqref{eq:stromgren_profile} is also shown, as well as the Str\"{o}mgren radius for both versions of the test.}
    \label{fig:stromgren_results}
\end{figure}

We discretize the medium on a $64^3$ cell grid, and perform the photoionization simulation using $10^6$ photon packets for 20 iterations. The results of both versions of the test are shown in \figureref{fig:stromgren_results} and are statistically equivalent to the results obtained with \cmacionizeone{} \citep{2018Vandenbroucke_CMacIonize}.

\begin{figure}[!h]
    \centering
    \includegraphics[width=0.48\textwidth]{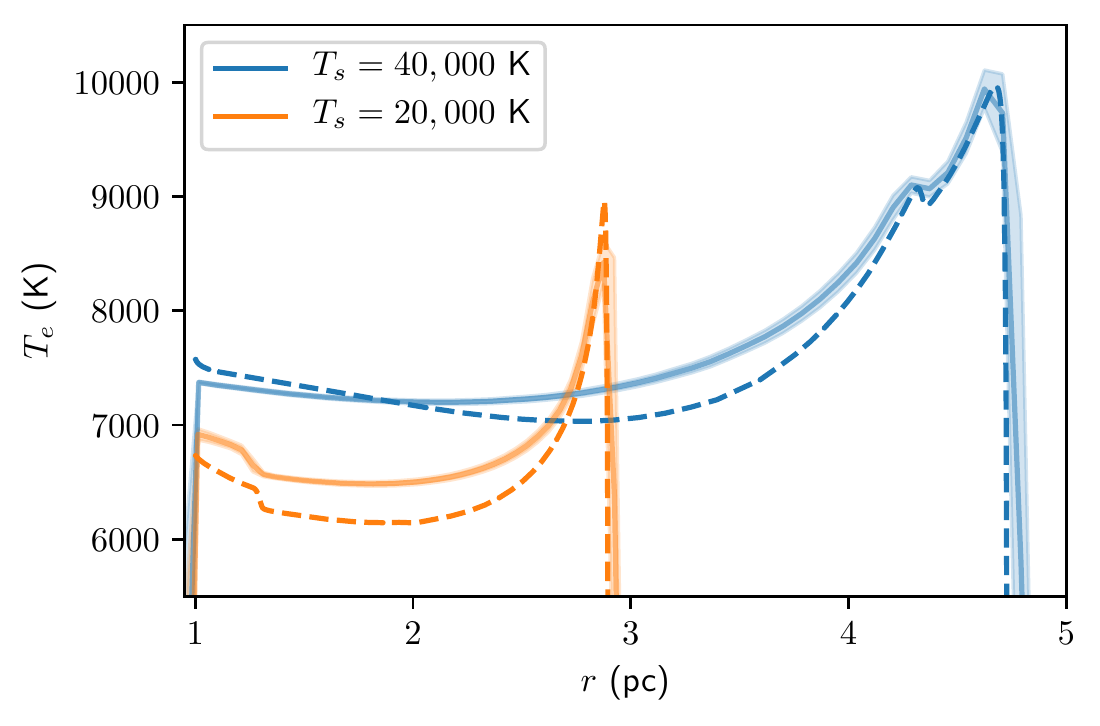}
    \caption{Temperature as a function of radius for the low and high temperature Lexington benchmarks. The simulation results were binned in 100 radial bins. The solid lines show the average value in each bin, while the shaded region underneath represents the standard deviation within each bin. The dashed lines correspond to the \textsc{Cloudy} reference solution.}
    \label{fig:lexington_temperature}
\end{figure}

\begin{figure*}[!h]
    \centering
    \includegraphics[width=0.98\textwidth]{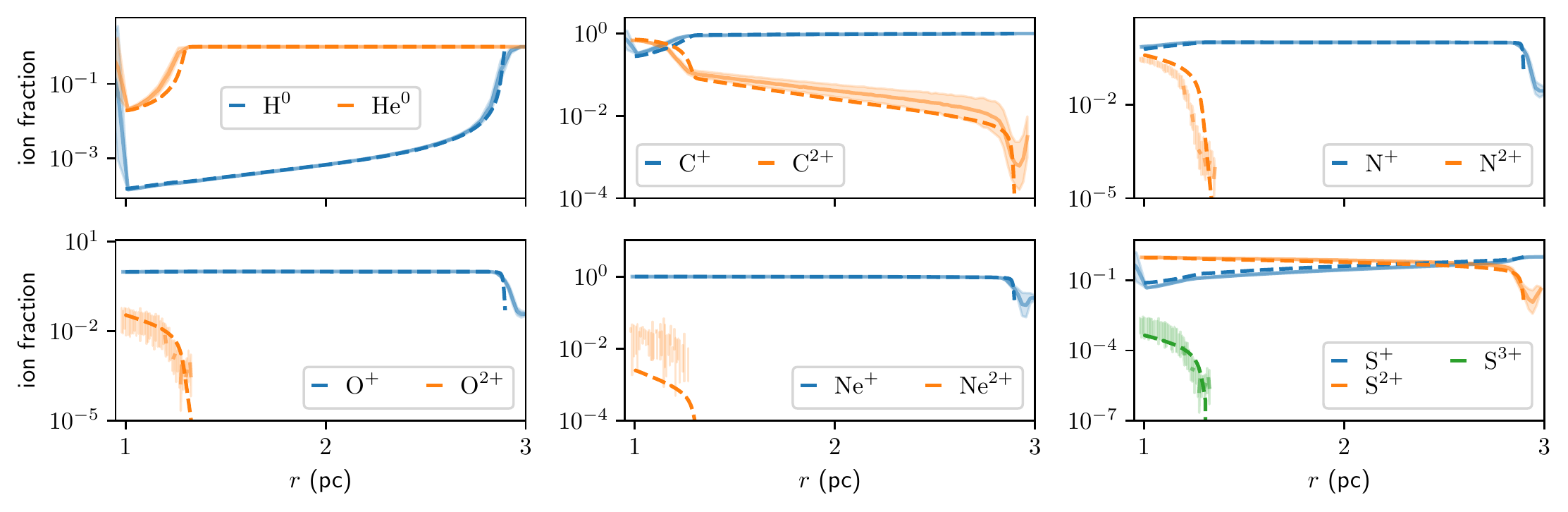}
    \includegraphics[width=0.98\textwidth]{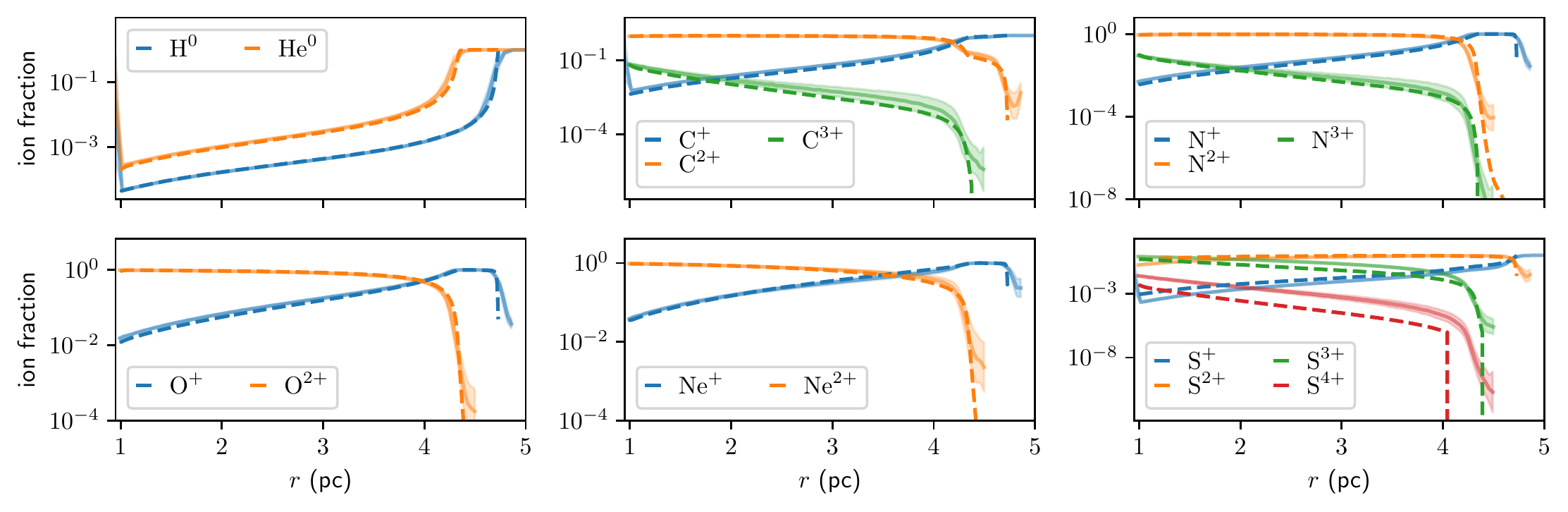}
    \caption{Ionic fraction as a function of radius for the low temperature (\emph{top rows}) and high temperature (\emph{bottom row}) Lexington benchmarks. The simulation results were binned in 100 radial bins in log space. The solid lines show the average value in each bin, while the shaded region underneath represents the standard deviation within each bin. The dashed lines correspond to the \textsc{Cloudy} reference solution.}
    \label{fig:lexington_ions}
\end{figure*}

\subsection{Lexington \textsc{Hii} region benchmark}

In an astrophysical photoionization (or \textsc{Hii}) region, a significant fraction of the UV extinction is due to the photoionization of helium. Furthermore, the photoionization cross sections for hydrogen, helium and other chemical elements depend on the energy of the ionizing photons, and the recombination rates depend on the local electron temperature. The temperature itself is an equilibrium value for which the photoionization heating is balanced by the radiative cooling due to deexcitation of excited states of predominantly metal ions. Some of the radiation released by recombining hydrogen and helium will still be emitted at ionizing frequencies, so that an additional diffuse ionizing field is generated. To properly resolve the structure of these regions, even for the basic case of a uniform medium with a single photoionizing source, it is hence necessary to track more chemical elements and to treat the photoionization cross sections, recombination rates, and temperature self-consistently.

Due to the complexity of this problem, a (semi-)analytic reference solution does not exist. However, a variety of benchmark tests have been formulated in literature, e.g. the Lexington benchmarks \citep{1995Ferland, 2001Pequignot}. These benchmarks consist of a $10\times{}10\times{}10$~pc uniform box with a spherical cavity with radius $1~$pc in the centre. In the centre of this cavity, a single black-body ionizing source is located. The difference between the two versions is the assumed temperature and ionizing luminosity of the source, $T_s=20,000~$K and $Q_{\rm{}H}=10^{49}~{\rm{}s}^{-1}$ for the low temperature benchmark, and $T_s=40,000~$K and $Q_{\rm{}H}=4.26\times{}10^{49}~{\rm{}s}^{-1}$ for the high temperature benchmark. The hydrogen number density outside the central cavity is set to $n_{\rm{}H}=100~{\rm{}cm}^{-3}$, and the density of the other elements we consider is given by the relative number abundances: ${\rm{}He}/{\rm{}H} = 0.1$, ${\rm{}C}/{\rm{}H}=2.2\times{}10^{-4}$, ${\rm{}N}/{\rm{}H}=4\times{}10^{-5}$, ${\rm{}O}/{\rm{}H}=3.3\times{}10^{-4}$, ${\rm{}Ne}/{\rm{}H}=5\times{}10^{-5}$ and ${\rm{}S}/{\rm{}H}=9\times{}10^{-6}$. All photoionization cross sections, recombination rates and re-emission probabilities are computed self-consistently using the same atomic data used by \citet{2018Vandenbroucke_CMacIonize} and are not part of the benchmark parameters.

In addition to testing the aspects of the basic photoionization algorithm that are also tested by the Str\"{o}mgren volume test, this test also tests the handling of diffuse photon packets, and the proper tracking and synchronisation of path length counters for elements other than hydrogen. It also tests the more elaborate temperature and ionization structure calculation at the end of each iteration (an aspect that has not changed from \cmacionizeone{}.

The benchmarks contained in the repository use a $64^3$ grid to discretize the medium, and require $10^8$ photon packets for 20 iterations. The number of photon packets required is significantly higher than in an equivalent Str\"{o}mgren volume test because of the low flux in the high energy tail of the black-body spectrum, which we need to sample accurately to get accurate ionic fractions for coolants with high ionization energies. $10^7$ photon packets are sufficient to reproduce the average temperature profile, while $10^8$ photon packets are required to get an acceptable noise level for the abundances of individual ions.

\figureref{fig:lexington_temperature} shows the temperature profiles for both versions of the test, as well as a 1D reference solution computed with \textsc{Cloudy} \citep{2017Ferland}, using the same input parameters. We find a reasonable agreement, with some deviations that can be traced back to differences between our physical model and that of \textsc{Cloudy}. Similar observations can be made for the ionic fraction profiles for both versions of the test, as presented in \figureref{fig:lexington_ions}. All results are again statistically equivalent to the results presented in \citet{2018Vandenbroucke_CMacIonize}, since nothing changed to our physical model.

\subsection{Turbulent boxes}

The two tests described above use a uniform medium and are essentially 1D tests because of the spherical symmetry of their setup. This is not representative for real applications, where the density structure and radiation field are truly 3D quantities that cannot be sampled using a 1D method. To validate our new algorithm in a 3D scenario, we need a setup with a range of densities. We can use \cmacionizeone{} to obtain reference results for the same setup.

 As 3D tests, we repeat the Str\"{o}mgren volume and high temperature Lexington benchmark tests, but this time using a turbulent ISM model as background density field. The turbulent field is generated by running \cmacionize{} in hydrodynamics only mode on a $256^3$ cell grid. Starting from a uniform medium, we drive turbulence using solenoidal forcing in Fourier space, with the method of \citet{1999Alvelius}. The simulation is evolved until the probability distribution function and power spectrum of the density and velocity fields reach a steady state. The box size and average density in the box are set to the same values as for the original test. Once the initial condition has been generated, the source is again put in the centre of the box. This time, we do not cut out a $1~$pc sphere for our alternative Lexington tests. Detailed results of these turbulent box simulations are described elsewhere (Sartorio \& Vandenbroucke, \emph{in prep.}).
 
 \begin{figure}
     \centering
     \includegraphics[width=0.48\textwidth]{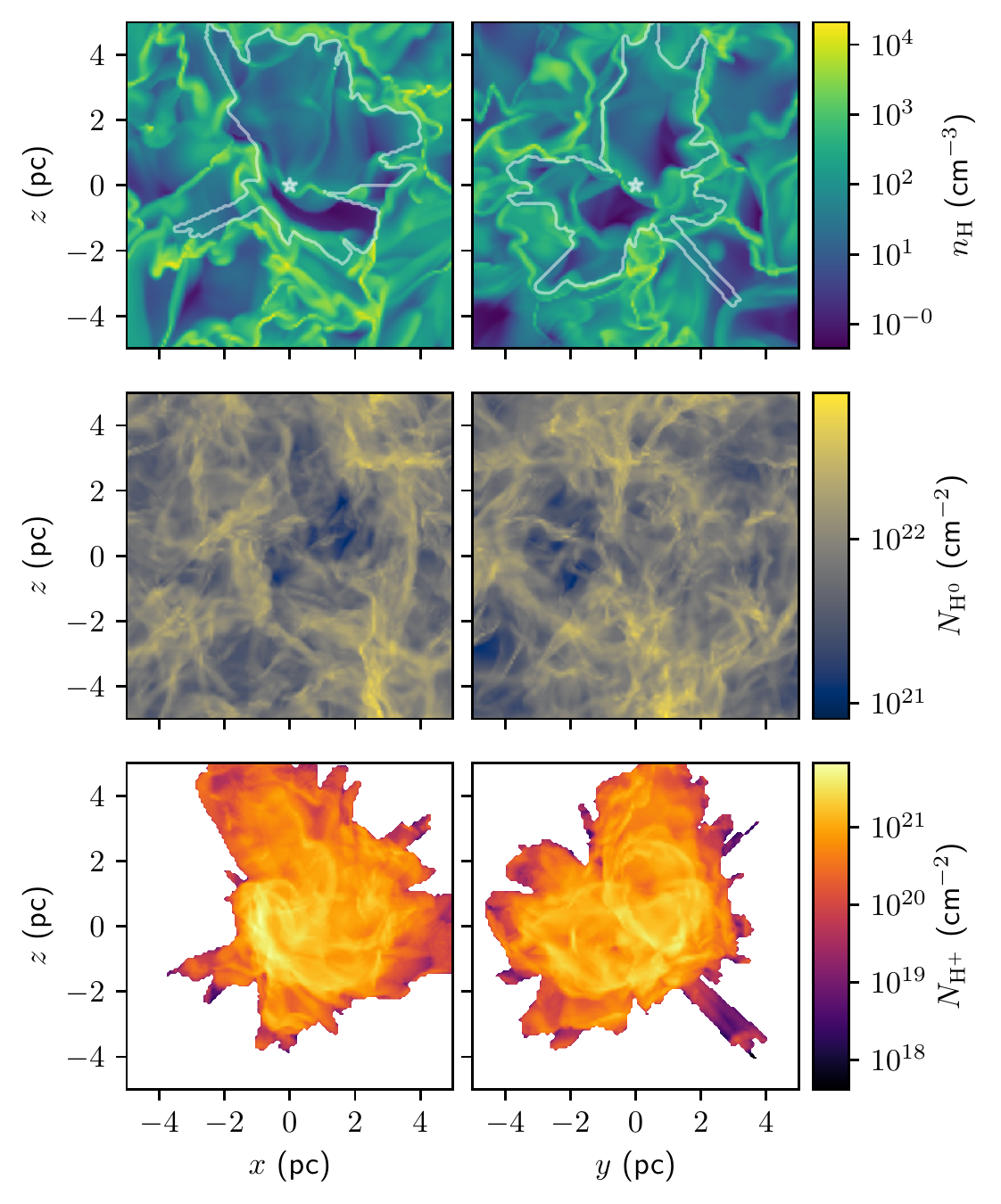}
     \caption{Str\"{o}mgren volume in a turbulent medium.  The columns correspond to different viewing angles: along the $y$ axis (\emph{left}) and along the $x$ axis (\emph{right}). \emph{Top row}: density in a planar cut through the centre of the box. The white star indicates the position of the ionizing source. The white contour shows the ionization front, defined as the radius where the neutral fraction rises above 0.5. \emph{Middle row}: surface density of the neutral gas. \emph{Bottom row}: surface density of the ionized gas. White patches correspond to sight lines along which all the gas is neutral.}
     \label{fig:turbulent_stromgren}
 \end{figure}
 
 \figureref{fig:turbulent_stromgren} shows the ionization structure of the turbulent Str\"{o}mgren volume test, after 10 iterations using $10^6$ photon packets. The density grid produced by the hydro simulation was resampled onto a $128^3$ grid to reduce the computational cost for this test. The ionizing source is located close to an intermediate density filament and ionizes a highly non-symmetric Str\"{o}mgren volume, predominantly in the directions perpendicular to the filament. The densities of the cells along the individual photon paths span three orders of magnitude. The results for \cmacionizeone{} and \cmacionize{} are indistinguishable. This can also be seen from the total ionized mass within the box: the old algorithm yields $266.18~{\rm{}M}_\odot{}$, while the new algorithm gives $266.13~{\rm{}M}_\odot{}$, consistent with Monte Carlo noise. Both these values are significantly lower than the $895.15~{\rm{}M}_\odot{}$ for a uniform medium, since the recombination rate is dominated by recombination in the filaments that is significantly higher than in the uniform case.
 
 \begin{figure}
     \centering
     \includegraphics[width=0.48\textwidth]{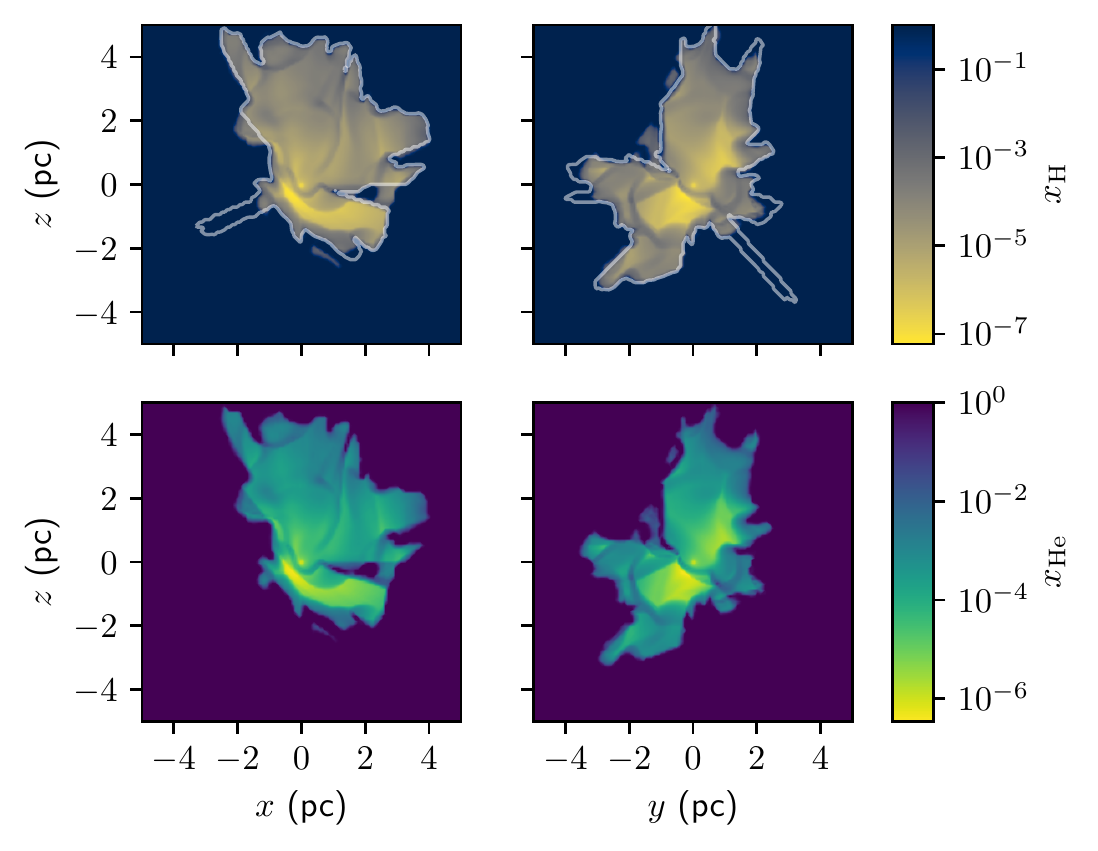}
     \caption{Hydrogen neutral fraction (\emph{top}) and helium neutral fraction (\emph{bottom}) in slices perpendicular to the $y$ axis (\emph{left}) and $x$ axis (\emph{right}), for the turbulent high temperature Lexington test. The white contour in the top panels shows the ionization front for the equivalent turbulent Str\"{o}mgren test, defined as the radius where the neutral fraction rises above $0.5$.}
     \label{fig:turbulent_H_comp}
 \end{figure}
 
 \begin{figure}
     \centering
     \includegraphics[width=0.48\textwidth]{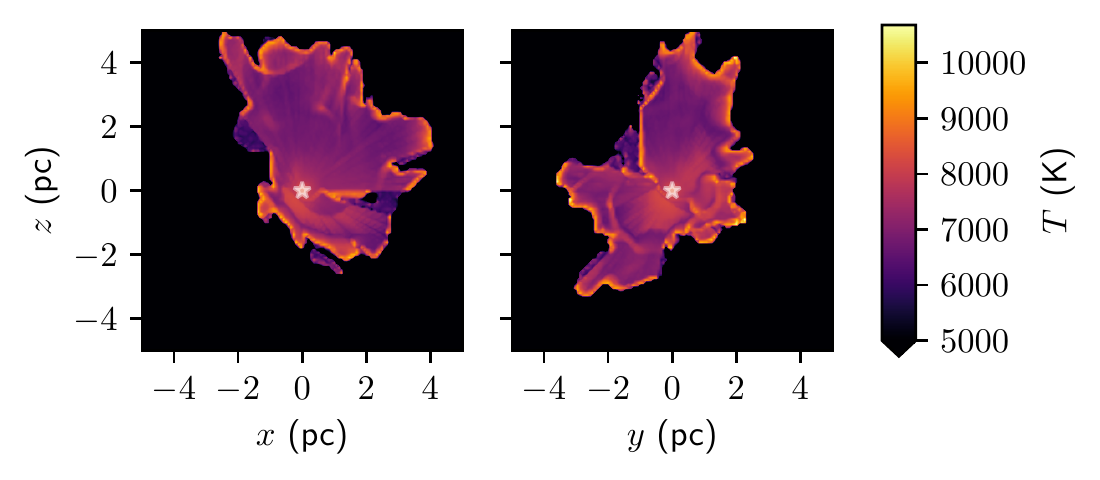}
     \caption{Temperature slices through the turbulent high temperature Lexington test. The two panels show slices perpendicular to the $y$ axis (\emph{left}) and perpendicular to the $x$ axis (\emph{right}). The white star indicates the position of the ionizing source. The temperature outside the \textsc{Hii} region is set to a fixed value of $500~$K and is therefore not included in the temperature scale.}
     \label{fig:turbulent_lexington_T}
 \end{figure}
 
 \begin{figure}[!h]
     \centering
     \includegraphics[width=0.48\textwidth]{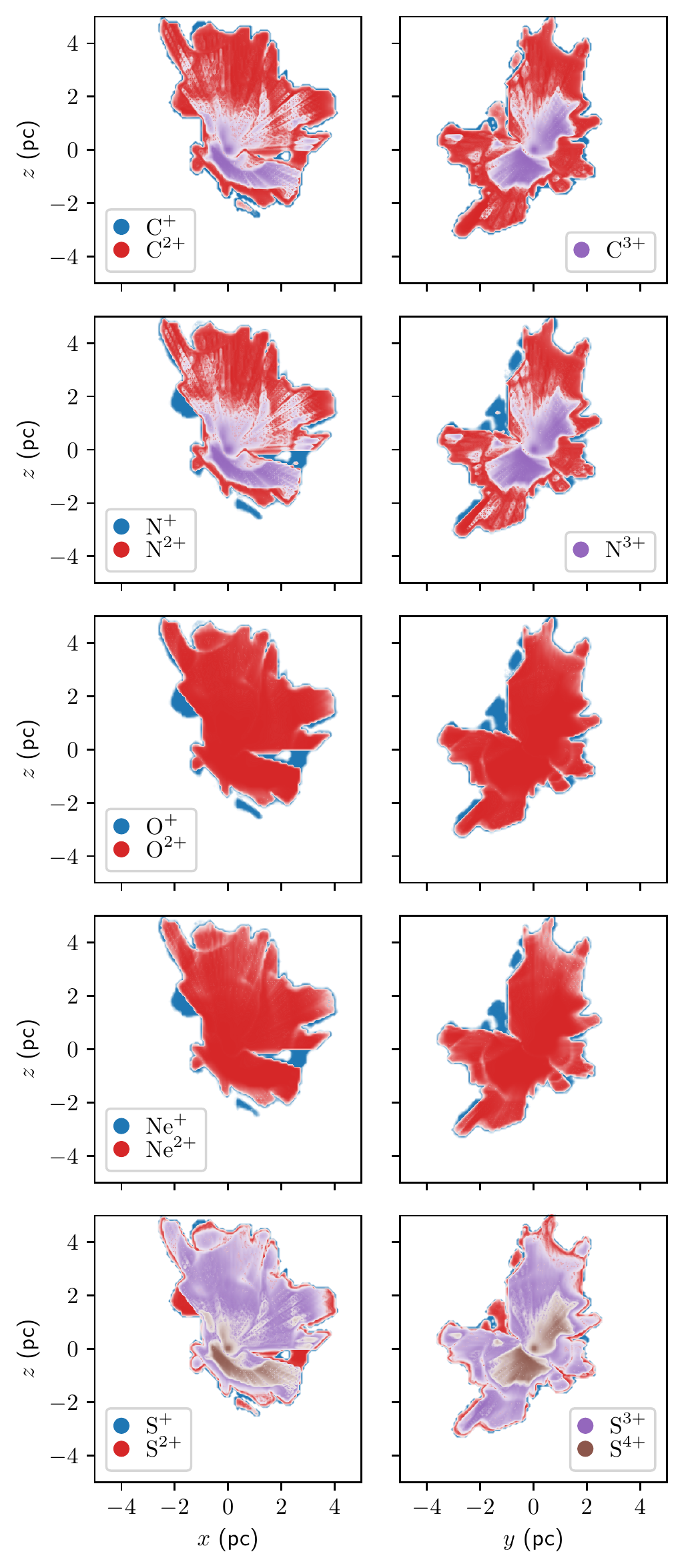}
     \caption{Ionic fractions of various ions in slices through the turbulent high temperature Lexington test. The two panels show slices perpendicular to the $y$ axis (\emph{left}) and perpendicular to the $x$ axis (\emph{right}). The opacity of the colours linearly scales with the ionic fraction of the corresponding ion. The ionic fractions outside the \textsc{Hii} regions are not accurately tracked by our method and are therefore not displayed.}
     \label{fig:turbulent_lexington_ions}
 \end{figure}
 
 \figureref{fig:turbulent_H_comp} shows the hydrogen and helium neutral fraction for the turbulent Lexington test. This test uses the same grid structure as the equivalent Str\"{o}mgren test, and uses $10^{8}$ photon packets for 20 iterations. As in the uniform Lexington test, the ionization fraction of helium is lower than that of hydrogen but follows the same general trend. The regions where hydrogen and helium are ionized almost have the same size, because the transition from ionized to neutral happens over a short distance near the denser filaments. The top panels of \figureref{fig:turbulent_H_comp} also show the hydrogen ionization front from the equivalent Str\"{o}mgren test for comparison. While in principle both tests should yield the same ionized region, the simplified physics treatment and the absence of a diffuse field in the Str\"{o}mgren test lead to small differences in the geometry of the \textsc{Hii} region.
 
 \figureref{fig:turbulent_lexington_T} and \figureref{fig:turbulent_lexington_ions} show the temperature and ionic fractions for the turbulent Lexington test. The general trends are the same as for the equivalent uniform test, shown in \figureref{fig:lexington_temperature} and \figureref{fig:lexington_ions}. The temperature is relatively constant and in the range $[7000,8000]~$K throughout most of the \textsc{Hii} region, and then steeply rises towards a temperature of $10,000~$K and more near the ionization front. The central parts of the \textsc{Hii} region are dominated by ions with high ionization energies, while ions with lower ionization energies start to dominate closer to the ionization front as absorption depletes high energy photons. Of significant interest are the regions that are not directly exposed to the star and hence can only have been ionized by diffuse radiation. One such region can clearly be identified to the right of the ionizing source in the $xz$ slices. This region is characterised by a relatively low temperature and dominated by ${\rm{}C}^{2+}$, ${\rm{}N}^{+}$, ${\rm{}O}^{+}$, ${\rm{}Ne}^{+}$ and ${\rm{}S}^{2+}$.

\subsection{Other tests}

The tests above verify all aspects of the new MCRT algorithm. \cmacionize{} has additional capabilities, such as the possibility to run radiation hydrodynamics simulations (see \sectionref{sec:RHD}), and these also have dedicated tests. We mention them here for completeness.

The coupling between photoionization and hydrodynamics is realised by assuming a two-temperature approximation, whereby hydrodynamical integration time steps are alternated with photoionization steps, and the hydrodynamic pressure is updated according to the ionization state returned by the MCRT algorithm. Neutral gas is set to a chosen low temperature, and ionized gas to a chosen high temperature. The temperature for cells that are partially ionized is obtained by linearly interpolating between these two temperatures. This approach is identical to that used by \citet{2015Bisbas}, and is tested using their \textsc{starbench} benchmark test that models the early expansion of a D-type ionization front. Results are consistent with \citet{2018Vandenbroucke_CMacIonize}.

\citet{2019Sartorio} extended the algorithm to also include the gravitational potential of a point source, in order to model trapped \textsc{Hii} regions surrounding massive protostars. To validate the coupling between hydrodynamics and the external potential, we created a new steady-state accretion flow test, similar to \citet{2019Vandenbroucke_Bondi}. This so-called Bondi test starts with a uniform medium and uses inflow boundary conditions that are consistent with the analytic Bondi profile for the assumed point mass potential at the radius of the boundaries. After a free-fall time scale, the density and velocity in the box will evolve into a steady-state solution that is consistent with the analytic Bondi profile. Since the Bondi profile diverges for zero radius, a central spherical region surrounding the point mass is masked out. The Bondi test hence tests the coupling between hydrodynamics and external gravity, the use of inflow boundary conditions, and the use of masked-out regions. In principle, the test could be extended to include a trapped \textsc{Hii} region, but since this setup is unstable \citep{2019Lund, 2019Vandenbroucke_Bondi}, this does not provide a practical benchmark test.

\section{Performance}
\label{sec:performance}

\subsection{Hardware specifications}

To test the performance of the new MCRT algorithm, we conducted a series of tests on the Tier-2 clusters of the VSC supercomputer at Ghent University. All tests were run on a single node with exclusive access. Since our new algorithm aims to benefit from efficient usage of memory caches, we selected two clusters with different memory cache characteristics: the low memory cluster \emph{golett} and the high memory cluster \emph{victini}.

The nodes of \emph{golett} consist of 2 CPUs that each have 12 cores, so a total of 24 cores. Each core has a 256~KB level 2 (L2) cache. The system has 4 non-uniform memory access (NUMA) domains with each a 15~MB level 3 (L3) cache. No hardware hyper-threads are available.

The nodes of \emph{victini} consist of 2 CPUs with 18 cores, so 36 in total. They have a significantly larger L2 cache of 1024~KB each. The system also has 4 NUMA domains, this time with L3 caches of 25~MB. Again, no hardware hyper-threads are available.

Since \emph{golett} has significantly less memory cache available than \emph{victini}, we expect to see differences in overall efficiency and parallel efficiency between the two clusters for different sizes of subgrids. This should cover a representative sample of contemporary hardware.

We have also run some tests on systems that support hyper-threading, to check if additional speedup can be achieved by exploiting hyper-threads. We find that this is generally not the case. Our algorithm is memory-bound and hence does not benefit from the availability of more computing power for the same memory bandwidth.

\subsection{Performance compared to a traditional MCRT algorithm}

Performance measurements are taken for two of the test cases discussed in \sectionref{sec:validation}: the hydrogen-only uniform Str\"{o}mgren volume test without diffuse field \citep{2018Vandenbroucke_CMacIonize}, and the 40000~K Lexington \textsc{Hii} region benchmark test \citep{1995Ferland, 2001Pequignot} that uses a more complex ISM model. For both tests, we use a grid with $128^3$ cells. The former test requires $10^6$ photon packets and 10 iterations, while the latter uses $10^7$ photon packets and 20 iterations. The number of photon packets for the latter test has been reduced relative to the tests described in \sectionref{sec:validation} to reduce the computational cost. For both tests, we run versions with a uniform and a turbulent density structure to cover a range of possible applications.

The Str\"{o}mgren test is challenging because of its lack of complexity. All photon packets are launched from a single source and the interaction between the photon packets and the medium is almost trivial. This means that the simulation is dominated by the overhead from the task-based algorithm, and that there is a high contention for the central subgrid that tests our subgrid copy algorithm. This is a representative scenario for many radiation hydrodynamics applications, where the coupling between the photoionization and the hydrodynamics is simplified to make problems tractable.

The Lexington test is a more representative test for the post-processing mode of the photoionization algorithm, where a high physical accuracy is more important than raw performance. In this test, the computational load is more evenly spread between the subgrids, and the subgrids themselves are considerably larger in memory, since we track a higher number of ions. This test challenges the memory management of our algorithm and should more easily expose performance gains caused by more efficient cache usage.

\begin{figure}
    \centering{}
    \includegraphics[width=0.48\textwidth{}]{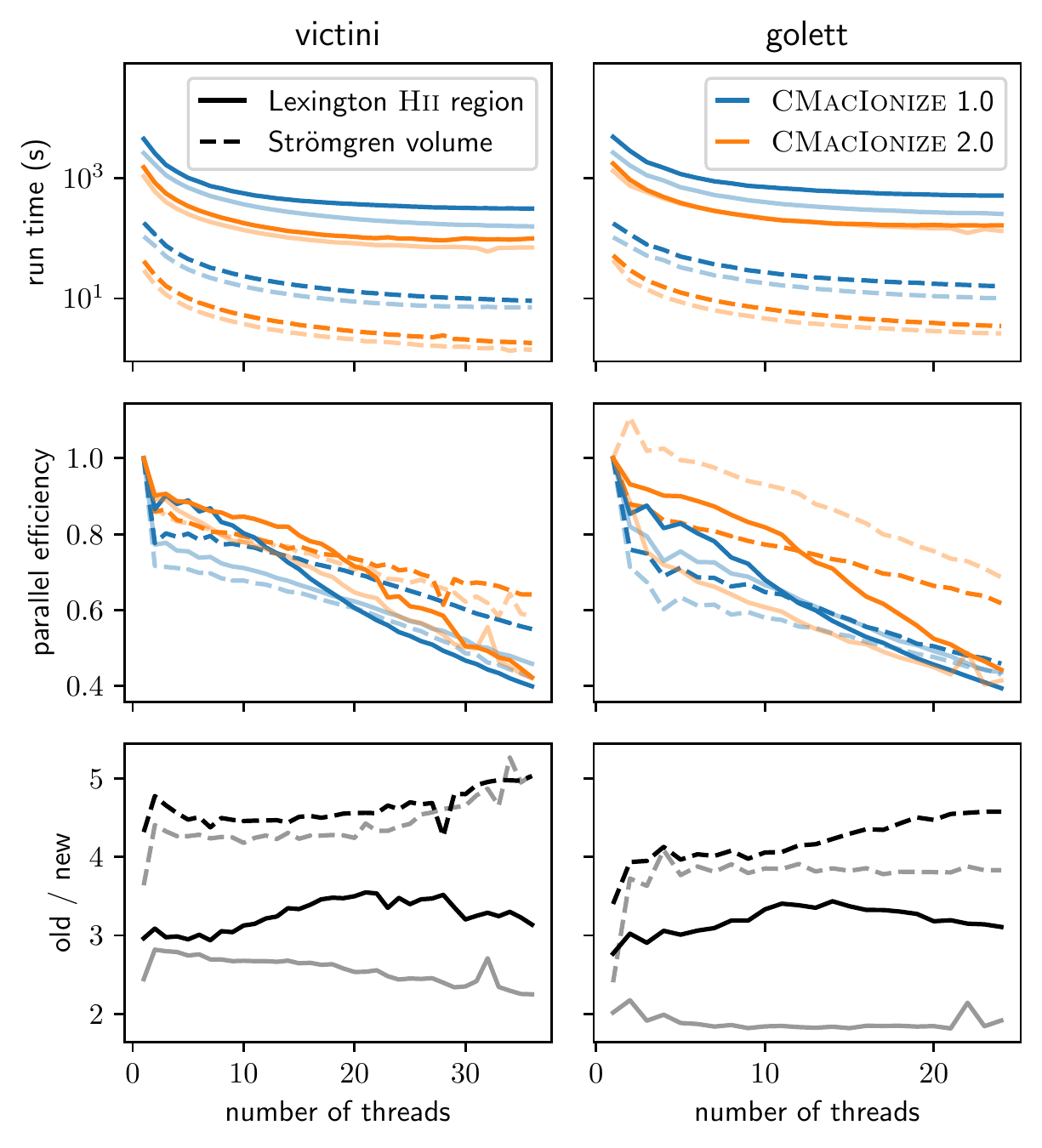}
    \caption{Total run time (\emph{top}), parallel efficiency (\emph{middle}), and algorithmic speedup (\emph{bottom}) for the Str\"{o}mgren and Lexington tests, run with two versions of \textsc{CMacIonize} using the same initial condition. The algorithmic speedup is computed by dividing the run time for the \cmacionizeone{} version by that for the \cmacionize{} version. The columns correspond to results for two different clusters. The dark lines correspond to the tests that use a uniform density structure, while the lighter lines are the same test in a turbulent density structure.}
    \label{fig:old_vs_new_time}
\end{figure}

\figureref{fig:old_vs_new_time} shows the total run time, parallel efficiency and algorithmic speedup for both tests, comparing \cmacionizeone{} and \cmacionize{}. All time measurements show the total accumulated time spent during photon propagation, and exclude the overhead due to initialisation and output, and the time spent in cell updates, which both versions of the algorithm have in common. We used $16^3$ cell subgrids for the Str\"{o}mgren test and $8^3$ cell subgrids for the Lexington test, corresponding to the optimal values found in \sectionref{sec:subgrid-size-tests}. The total run time for the two tests is considerably lower when using the task-based algorithm, with an algorithmic speedup of around 2 for the Lexington test, and around 4 for the Str\"{o}mgren test. There is also a significant improvement in parallel efficiency, especially for the node with small caches and for the Lexington test that is more memory-intensive.

The algorithmic speedup is more significant for the uniform than for the turbulent density structures. This is mainly caused by a reduction in the total time spent doing photon propagation, since the ionized region is smaller in the non-uniform case. As a result, relatively more time is lost due to the inevitable load imbalances at the end of each iteration. We expect the algorithmic speedup to get better for higher numbers of photon packets, when the total time for each iteration increases and load imbalances become relatively smaller.

\subsection{Subgrid size}
\label{sec:subgrid-size-tests}

The most significant parameter that determines the performance of the new task-based algorithm is the size of an individual subgrid. This parameter is important for two reasons: it determines the number of subgrids and with that the granularity of the tasks, and it determines the memory requirements for the propagation tasks that constitute the bulk of the run time. The best performance can be expected to be a trade-off between subgrids that are small enough to fit in fast memory cache, but large enough to limit the number of tasks and the associated overhead.

To test the impact of the subgrid size, we run full scaling tests for both our tests using
subgrid sizes of $4^3, 8^3, 16^3$ and $32^3$ cells. The associated memory sizes for these subgrids are shown in \tableref{table:struct_sizes}. All tests use a source copy level parameter of 4, which results in a total of respectively 32917, 4245, 661 and 186 subgrids.

To analyse the performance, we focus on the execution time of the various tasks as output by the code at run time. Each task records the start and end time of its execution, and each thread accumulates these time intervals per task. For each iteration, we then get a full overview of how much time each thread spent in each task. Additionally, we also record the start and end time of the iteration separately, to get the total elapsed time for the iteration in the same units.  All times are measured using the ticks of the internal clock of the CPU corresponding to $\approx{}0.40$~ns on \emph{golett} and $\approx{}0.44$~ns on \emph{victini}.

The difference between the total iteration time and the sum of all task execution times for a thread yields the idle time for that thread. This idle time is caused by the overhead of managing the tasks and by load imbalances. It is not possible to distinguish between these two causes based on the run time diagnostics without affecting the overall performance.

\begin{figure}
    \centering{}
    \includegraphics[width=0.48\textwidth{}]{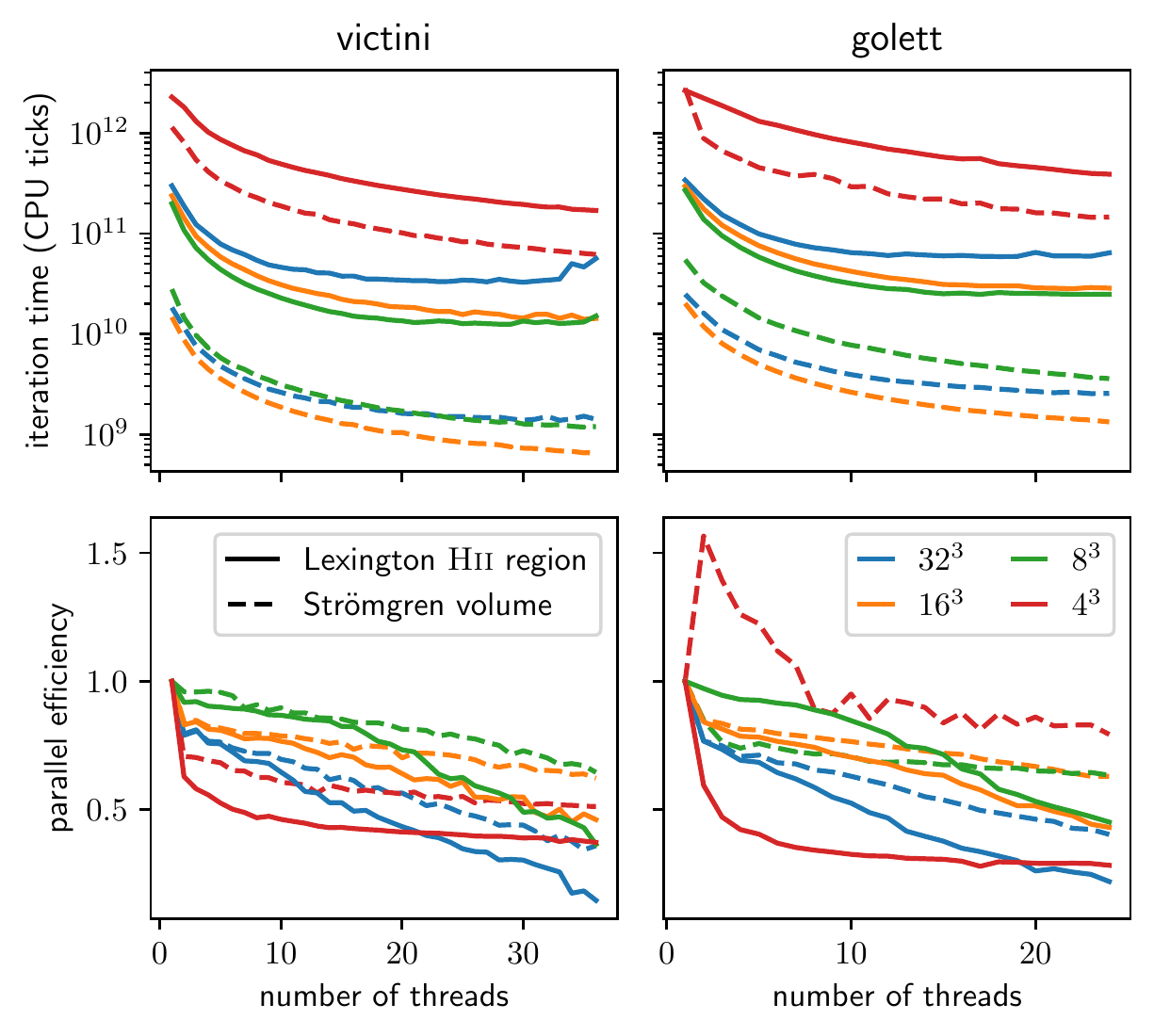}
    \caption{Total run time (\emph{top}) and parallel efficiency (\emph{bottom}) for a    single iteration of the Str\"{o}mgren and Lexington tests, as a function of the number of shared memory threads used to execute the test. The columns show the results for two different clusters, while the colours correspond to different subgrid sizes, as indicated in the legend.}
    \label{fig:nsubgrid_time}
\end{figure}

\figureref{fig:nsubgrid_time} shows the total iteration time and the corresponding parallel efficiency for all runs. It is clear that the subgrid size has a significant impact on the total run time, with the longest run time being recorded for the smallest subgrids. The parallel efficiency fluctuates strongly and is even super-optimal for one of the tests, illustrating the significant impact of changes in memory usage on parallel performance. The best scaling is observed for tests that use intermediate size subgrids, and for the Str\"{o}mgren test that requires less memory per cell.

\begin{figure}
    \centering{}
    \includegraphics[width=0.48\textwidth{}]{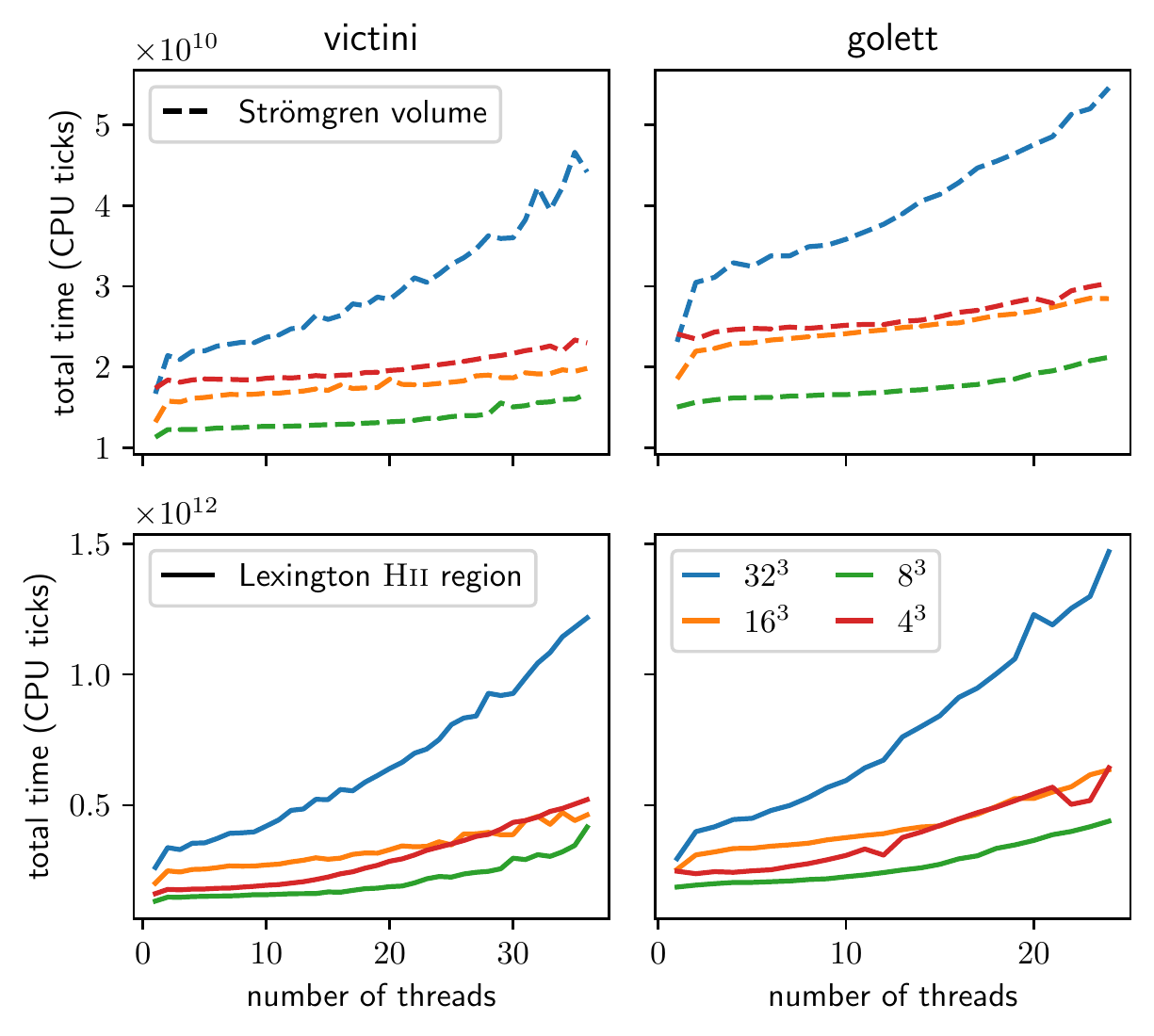}
    \caption{Total time spent in propagation tasks during the tests shown in \figureref{fig:nsubgrid_time}.}
    \label{fig:nsubgrid_task_times}
\end{figure}

To explain these differences, it is instructive to look at the accumulated time for the propagation tasks, shown in \figureref{fig:nsubgrid_task_times}. Each of these tasks consists of a payload of computational work, and a small wrapper of overhead required to start and finish the task. The work done as part of the payload is independent of our subdivision of the grid, so the sum of all payloads should be the same, no matter what the subgrid layout is. In an ideal scenario, the task execution is dominated by the payload, and the accumulated times should be independent of the subgrid layout as well.

If execution of the payload only depended on the speed of the CPU, then in reality the total sum of all task executions would be an increasing function of the number of tasks, since the overhead would constitute an increasing fraction of the task time. This is the behaviour we expect for the propagation tasks when the number of subgrids increases. However, it can be clearly seen that the total time spent in propagation tasks is lower for the runs with $8^3$ and $16^3$ cell subgrids than for the run with $32^3$ cell subgrids, despite these runs having more subgrids and hence more propagation tasks. Only for the $4^3$ cell subgrid run with the smallest subgrids do we observe the expected increase in total time due to overhead, and only for the Str\"{o}mgren test where the payload is smaller and the overhead makes up a larger fraction of the task.

Another observation is that the total time spent in propagation tasks is only a weak function of the number of threads for most tests, except for the test with the largest subgrids. This can be explained by comparing the memory sizes of the subgrids with the sizes of the L2 and L3 caches on the test machines. The largest subgrid in all cases has a size that is significantly larger than the L2 cache, meaning that it can only benefit from the slower L3 cache. The $8^3$ cell subgrid on the other hand fits in the L2 cache, allowing the propagation task to benefit from much faster memory access. The $16^3$ cell subgrid fits in the L2 cache on the \emph{victini} node for the Str\"{o}mgren test, but not for the Lexington test, while it does not fit for either test on the \emph{golett} node. On top of that, the significantly larger L3 cache on \emph{victini} enables more subgrids to be stored close to the CPU in between tasks, a feature that we actively exploit in our task scheduling strategy.

\begin{figure}
    \centering{}
    \includegraphics[width=0.48\textwidth{}]{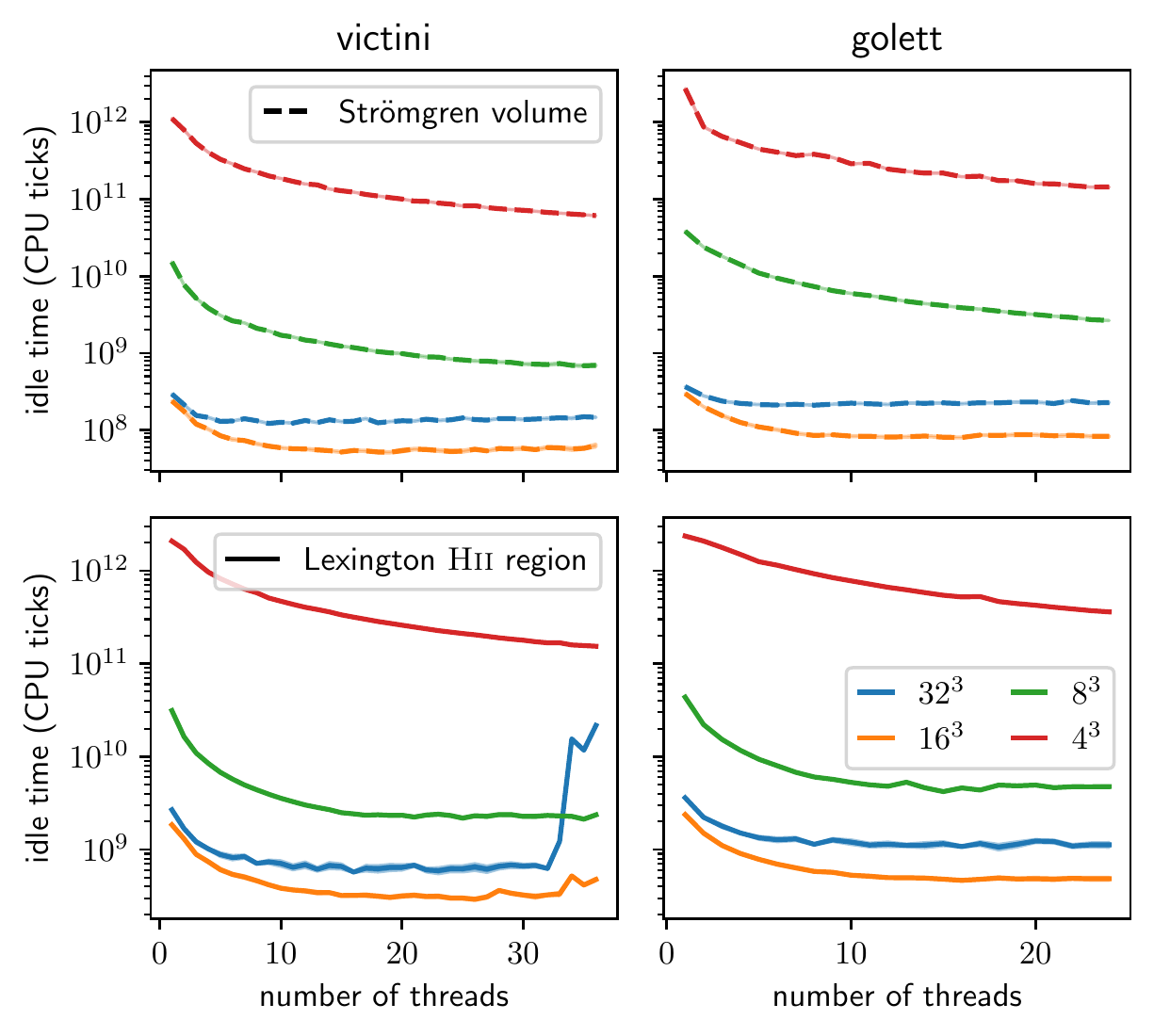}
    \caption{Average idle time per thread as a function of number of threads for the tests shown in \figureref{fig:nsubgrid_time}. The spread around the average for the different threads is shown by the shaded regions and is fairly small.}
    \label{fig:nsubgrid_idle_time}
\end{figure}

Another significant factor affecting the overall performance is the idle time. The average idle time per thread and the spread among the threads is shown in \figureref{fig:nsubgrid_idle_time}. The average idle time can be considered a proxy for the overhead associated with the task management, while the spread could be a proxy for load imbalances, although it is important to stress that we cannot actually disentangle these factors with the limited diagnostics from these runs. It is clear that the overhead increases very significantly with the number of subgrids, to the point that it completely dominates the run time for the tests with the smallest subgrids. The spread is fairly small for all tests, and is only visible in the figure for the Lexington tests with the largest subgrids. This is expected, since for this test the low number of subgrids makes load balancing harder to achieve. Bad load balancing will also trigger more premature buffer launches, which in turn leads to the creation of more small propagation tasks that generate more idle time. This effect is clearly visible in the bottom left panel of \figureref{fig:nsubgrid_idle_time}, where the average idle time significantly increases for high thread numbers.

\subsection{Subgrid copies}

To show the importance of subgrid copies, we compare two versions of the Str\"{o}mgren test: a version without any subgrid copies and a version that uses a source copy level of 4 (identical to the tests in \sectionref{sec:subgrid-size-tests}). We focus on a test with $16^3$ cell subgrids on \emph{victini}.

\begin{figure}
    \centering{}
    \includegraphics[width=0.48\textwidth{}]{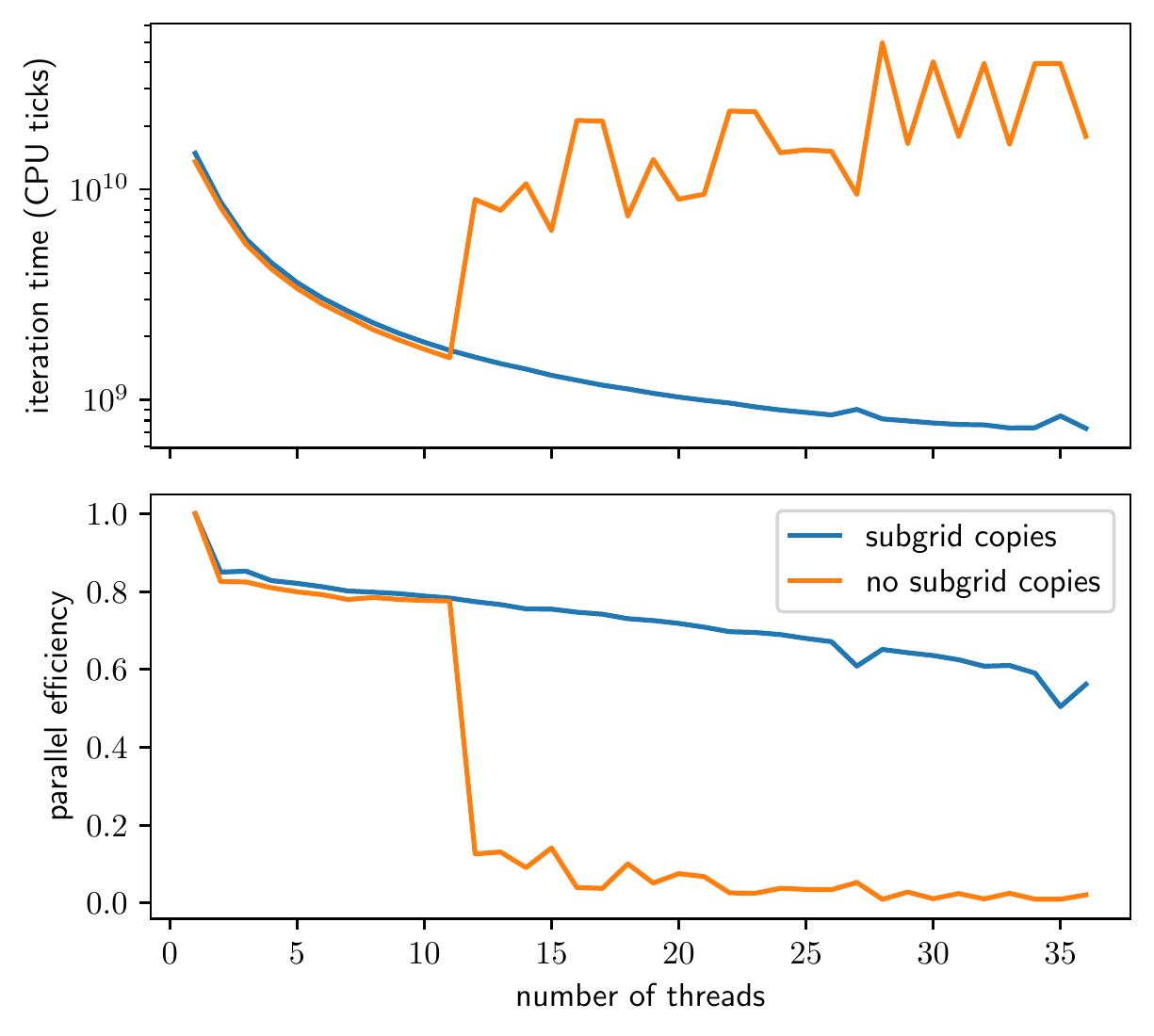}
    \caption{Total iteration time (\emph{top}) and parallel efficiency (\emph{bottom}) for two runs with different values of the source copy level parameter.}
    \label{fig:copies_scaling}
\end{figure}

\figureref{fig:copies_scaling} shows the total iteration time and parallel efficiency for both versions of the test. It is clear that the absence of subgrid copies leads to a significant loss of parallel efficiency for high thread numbers, while the presence of subgrid copies does not significantly affect the overall run time at low thread numbers.

\begin{figure*}
    \centering{}
    \includegraphics[width=0.98\textwidth{}]{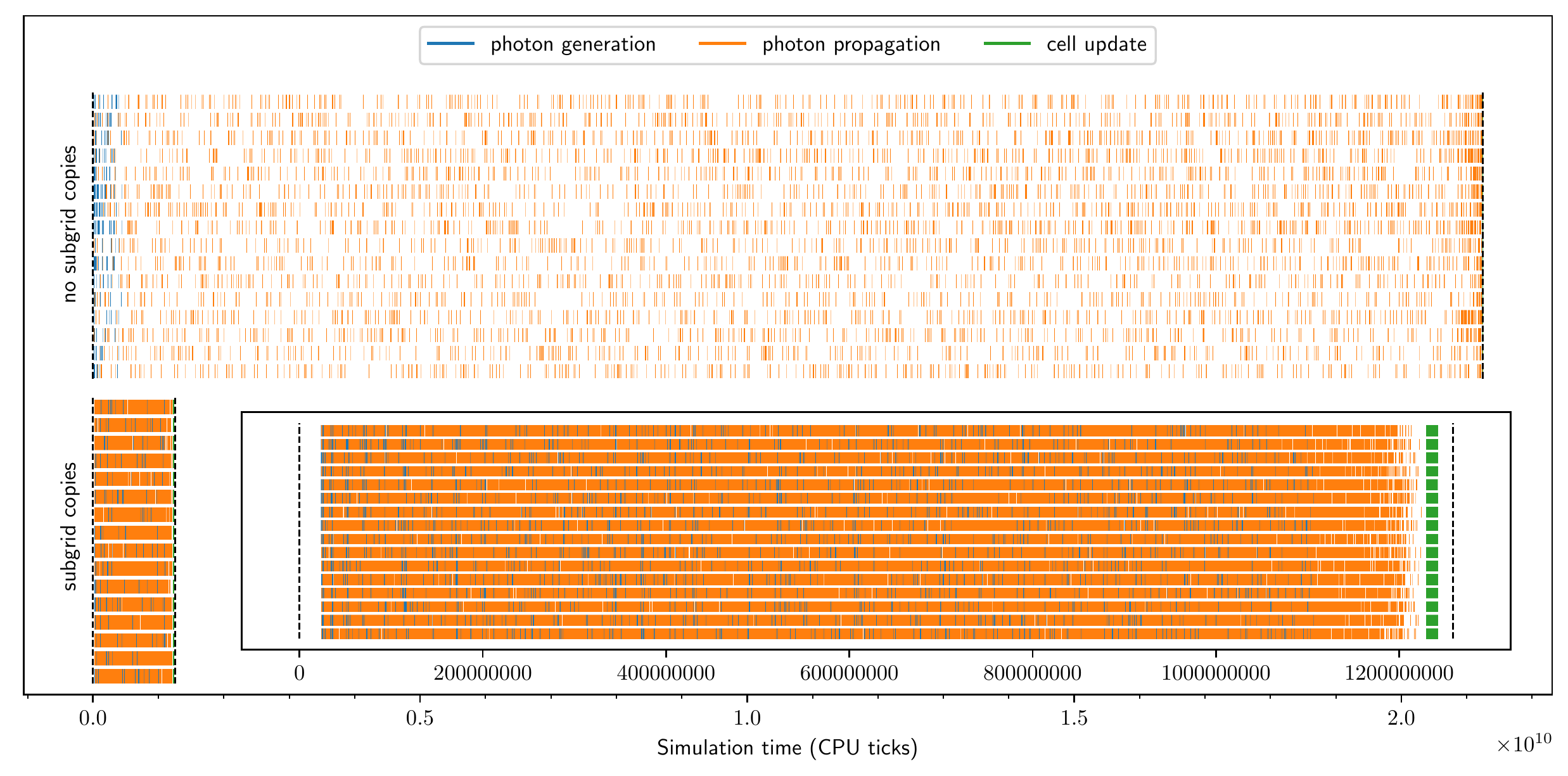}
    \caption{Task execution time line for a 16 thread run with and without subgrid copies. Each horizontal line corresponds to a single thread. The colours correspond to the different tasks that are being executed by the thread, as indicated in the legend. White segments correspond to idle time. The dashed lines at the start and end correspond to the start and end of the iteration. The inset shows a zoom for the run with subgrid copies.}
    \label{fig:copies_taskplot}
\end{figure*}

\begin{figure}
    \centering{}
    \includegraphics[width=0.48\textwidth{}]{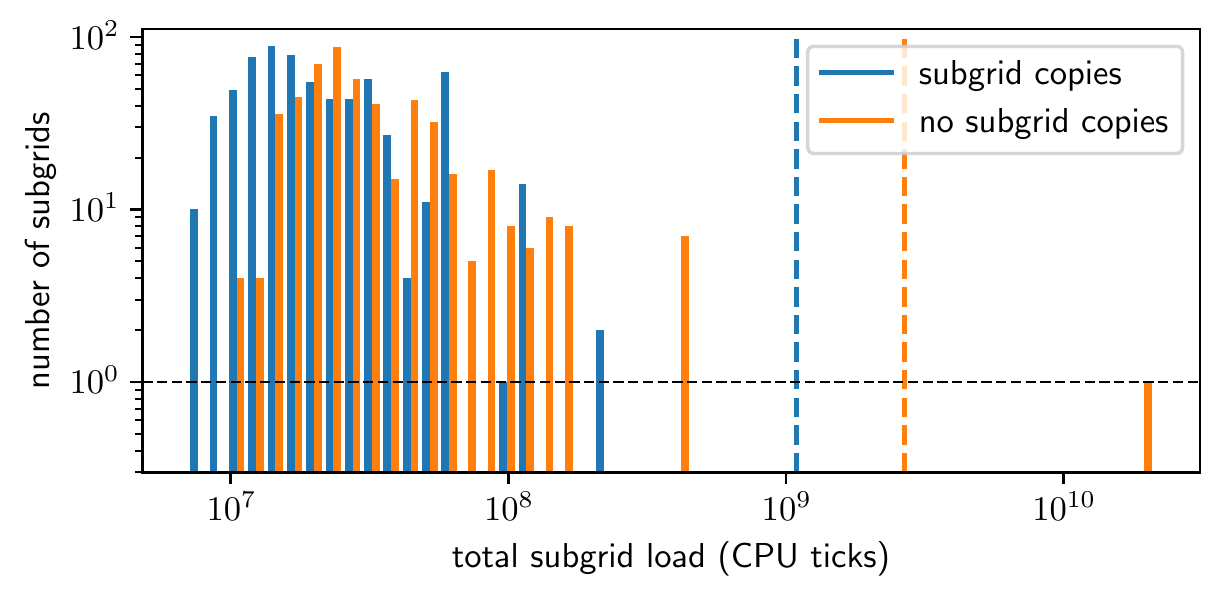}
    \caption{Histograms showing the total propagation task load for all subgrids for the tests with and without subgrid copies. The horizontal dashed line indicates a single subgrid in a bin for clarity. The vertical dashed lines represent the total subgrid load divided by the number of threads, acting as a proxy for the average load capacity per thread.}
    \label{fig:copies_histogram}
\end{figure}

We can explore the impact of copies by looking at a visual representation of the task execution time line for both versions of the test. This can be achieved by recording the start and end time of each task during task execution, and writing out these values at the end of the run. This causes a small overhead during task execution and means we cannot reuse task data structures, so that this technique is only practical for small runs. The resulting \emph{task plot} is shown in \figureref{fig:copies_taskplot} for a test with 16 threads. The version without subgrid copies shows a significantly higher fraction of white gaps caused by load imbalances.

These load imbalances are caused by the relatively high cost of the propagation tasks for the central subgrid that hosts the radiation source, as is shown in \figureref{fig:copies_histogram}. Since the fraction of the total computation time spent in this central subgrid is higher than the average computation capacity per thread, it is impossible to schedule the tasks in a favourable way. By making copies of the central subgrid, the total cost of this subgrid is divided among multiple subgrids, and good load balancing is restored. The improved cache efficiency that is caused by better load-balancing in turn reduces the average load per subgrid, as is also visible in the histogram.

The presence of subgrid copies is also beneficial if the central subgrid does not have a significantly higher load, but is still accessed considerably more than average, as e.g.\ for the Lexington test. Even in this case, we do notice a considerable impact of using subgrid copies for high thread numbers, because the copies reduce contention for the locks of the central subgrid.

\subsection{Memory usage}

\begin{figure}
    \centering{}
    \includegraphics[width=0.48\textwidth{}]{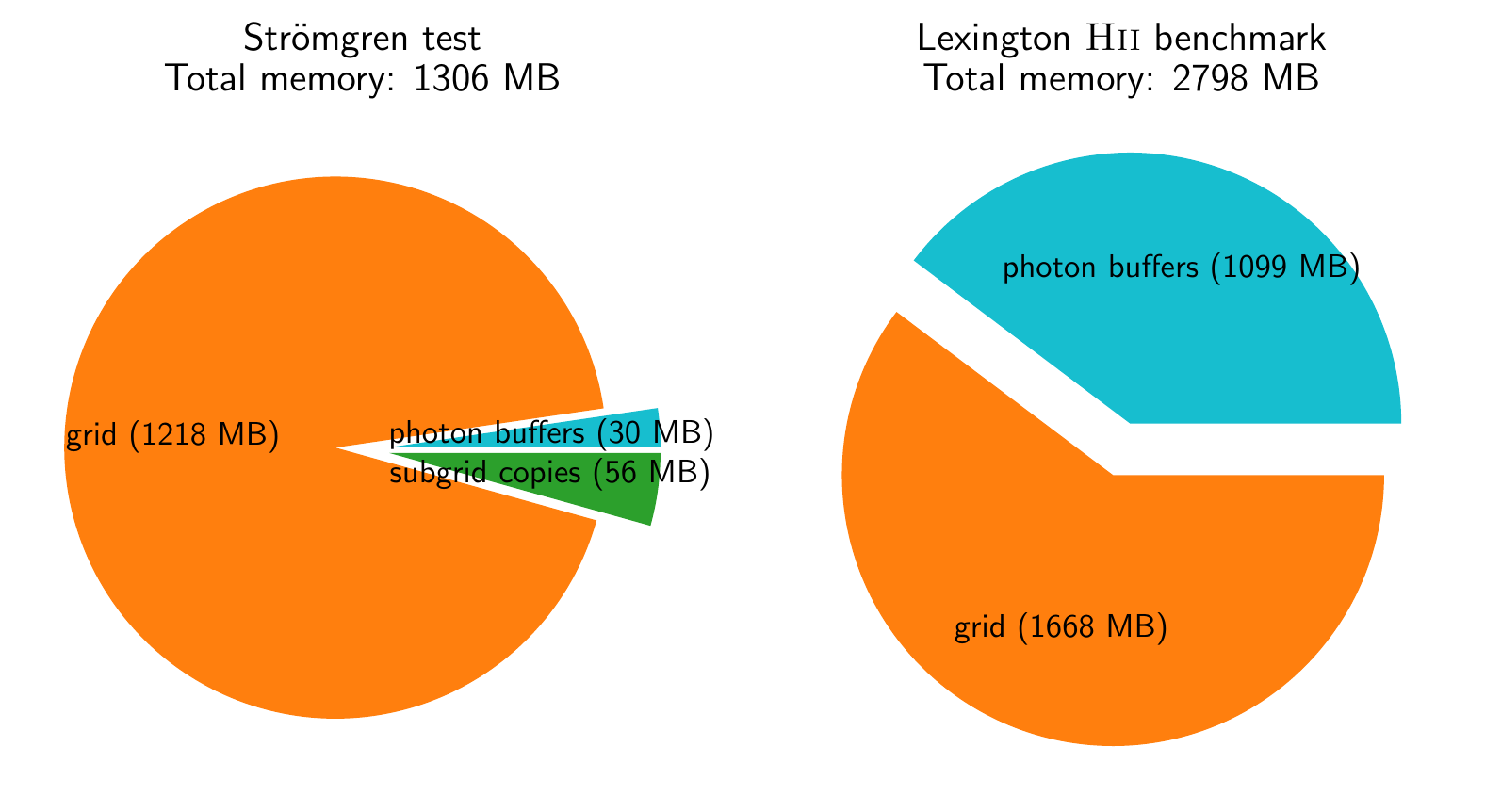}
    \caption{Memory requirements of the various data structures measured for two tests using 16 threads on \emph{victini}. Only contributions of more than one per cent of the total are shown.}
    \label{fig:memory_piechart}
\end{figure}

The task-based MCRT algorithm inevitably has a larger memory footprint than a traditional MCRT algorithm because of the need to store a considerable number of photon packets during the simulation. Subgrid copies cause additional memory overhead. Apart from the grid, which still dominates memory usage for all practical applications, the photon packet buffers represent the largest contribution to the memory footprint of the algorithm, as illustrated in \figureref{fig:memory_piechart}. It is therefore essential to understand what determines the required number of photon buffers and set that parameter to an appropriate value for each application.

The number of buffers in use during a simulation varies over time, as new buffers are activated by generation and propagation tasks. In the worst case scenario, every subgrid and every subgrid copy require seven active buffers: one for each neighbouring subgrid and an additional buffer to store photon packets that have been absorbed locally. In practice, not all these buffers will necessarily be activated, since some parts of the grid could receive no or very little photon packets. Moreover, each queued propagation task requires a buffer that is not linked to any subgrid. The number of queued tasks increases with the number of queues (and hence with the number of threads) and in the worst case scenario can be predicted from the maximum number of tasks in each queue. In runs with diffuse re-emission, scheduled re-emission tasks in the shared queue also contribute to the buffer requirements. In order to estimate the memory usage of the buffers, we need to measure the maximum queue load, and explore how it varies for different numbers of threads, different numbers of subgrids, and different MCRT parameters.

\begin{figure*}
    \centering{}
    \includegraphics[width=0.98\textwidth]{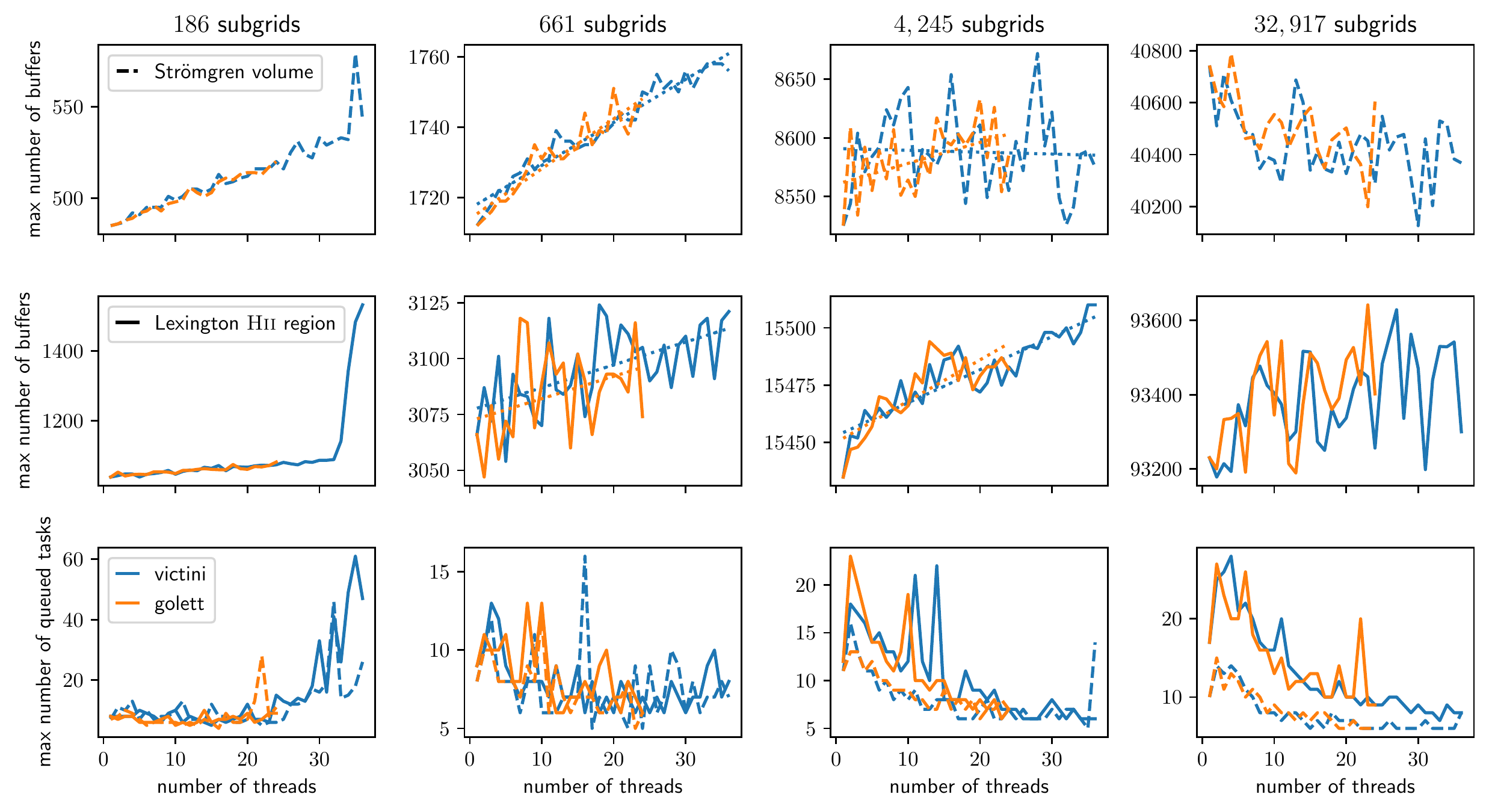}
    \caption{Maximum number of active buffers for the Str\"{o}mgren (\emph{top}) and Lexington (\emph{middle}) tests, and maximum number of queued tasks in any per-thread queue (\emph{bottom}) as a function of number of threads for the two different clusters. The columns correspond to the four tests with different subgrid sizes shown in \figureref{fig:nsubgrid_time}. The dotted lines in the central panels indicate a linear fit to the corresponding curves.}
    \label{fig:memory_maxima}
\end{figure*}

The maximum number of tasks among all per-thread queues and the maximum number of active buffers for our tests is shown in \figureref{fig:memory_maxima}. As expected, the number of active buffers is a strong function of the number of subgrids, supplemented with an apparently linear dependence on the number of threads. The maximum number of tasks in any of the thread queues is approximately independent of the number of threads, or in other words, the queue load is approximately linear in the number of threads as well. Overall, the number of tasks that is waiting in a queue is low due to our adopted task execution strategy.

There are two notable exceptions to these observations. The Lexington test with a low number of subgrids shows a sharp increase in buffers and queued tasks for high thread numbers. This is caused by the load imbalances we noted before, which lead to the creation of an excessively high number of prematurely scheduled propagation tasks. The Str\"{o}mgren test for a high number of subgrids does not show a linear dependence of the number of buffers on the number of threads, linked to the observed decrease in queue load with thread number. This indicates that tasks are executed very efficiently in this case; recall that these runs are dominated by overhead.

To obtain a practical model for predicting the memory requirements for a run, we focus on the two intermediate cases (with $8^3$ and $16^3$ cell subgrids), and fit a linear function to the number of buffers. These fits are also shown in \figureref{fig:memory_maxima}. If we write the linear function as
\begin{equation}
    n_{\rm{}buf} = A n_{\rm{}subgrid} + B n_{\rm{}thread}, \label{eq:heuristic}
\end{equation}
then all fits have values for $A$ in the range $[2.0, 2.6]$ for the Str\"{o}mgren test and in the range $[3.6, 4.7]$ for the Lexington test. This means that in practice only about half of the subgrid buffers are active at the same time. Except for the clearly aberrant fit, the $B$ coefficients all have values in the range $[1,2]$, confirming that the average queue load is indeed quite low.

We conclude that the function \eqref{eq:heuristic} with $A=5$ and $B=2$ provides a conservative practical estimate for the number of buffers, provided that the selected subgrid size results in a small load imbalance and idle time.

\begin{table}
    \centering
    \caption{Size in memory for photon packet buffers, individual cells and subgrids of various sizes for the two modes of the photoionization algorithm.}
    \begin{tabular}{ c c c }
    \hline
    Data structure & Hydrogen-only mode & Full mode \\
     & (Str\"{o}mgren test) & (Lexington test) \\
    \hline
    photon buffer & 17.2 KB & 37.52 KB \\
    single cell & 96 bytes & 310 bytes \\
    $4^3$ cell subgrid & 6.46 KB & 20.0 KB \\
    $8^3$ cell subgrid & 48.5 KB & 156 KB \\
    $16^3$ cell subgrid & 384 KB & 1.22 MB \\
    $32^3$ cell subgrid & 3.00 MB & 9.75 MB \\
    \hline
    \end{tabular}
    \label{table:struct_sizes}
\end{table}

For convenience, the size of the individual components that dominate the memory load is shown in \tableref{table:struct_sizes} for the two main MCRT modes supported by \cmacionize{}: the hydrogen-only mode used by the Str\"{o}mgren test and the full physics mode used by the Lexington test. Note that the code needs to be configured explicitly to use the hydrogen-only mode that uses less memory, and that these sizes are likely to change when additional physics are added to the code.

\section{Extensions of the algorithm}
\label{sec:extensions}

\subsection{Extended sources}
\label{sec:smooth_sources}

The algorithm as discussed so far has assumed that photon packets are emitted by discrete sources with a well-defined position. However, this need not be the case. Radiation could be incoming from an external radiation field such as the cosmic UV background, or could be emitted by a smooth source such as a stellar surface or a galactic luminosity distribution. For these extended sources, the photon packet generation task does not only provide the packet's initial energy and propagation direction, but also draws a random starting position from the source's spatial distribution. This position can be anywhere in the simulation box, so that an extended source cannot be associated with a particular subgrid. Therefore, a photon packet generation task for an extended source uses multiple output photon buffers, one for each subgrid in the simulation. For every newly created photon packet, the task locates the subgrid containing the starting position and adds the packet to the corresponding buffer.

After all photon packets for an extended source have been created, output buffers that are only partially filled  still contain photon packets that have not been scheduled. A separate task of a new type is then created to \emph{flush} these buffers. This task performs a simple iteration over all extended source output buffers and creates a photon packet propagation task for each one that is non-empty.

The photon packet generation tasks for a given extended source need exclusive access to all of the source's output buffers. As a result, only one such task can be executed at any given time. In a simulation where a particular  extended source acts as the dominant or only source of photon packets, this bottleneck can have a significant effect on the overall parallel efficiency. The problem can be overcome by providing multiple sets of the extended source photon packet output buffers, effectively treating the dominant source as multiple lower-luminosity sources.

\subsection{Dust radiative transfer}
\label{sec:dustRT}

The algorithm presented in \sectionref{sec:algorithm} and implemented in \cmacionize{} addresses photoionization simulations. The technique could be adjusted or extended for other MCRT applications in a straightforward fashion. In this section we consider some of the issues related to performing dust MCRT using a variation of our algorithm, without actually developing an implementation. The interactions between the radiation and the medium now include absorption, scattering and thermal re-emission (at longer wavelengths). 

\paragraph{Radiation field} Tracking the local radiation field as a function of wavelength \citep{1999_Lucy,2003_Niccolini} is a trivial extension of our mechanism for tracking the ionisation state of the medium. Because multiple values need to be stored per cell (one per wavelength bin), the memory footprint of a subgrid increases, possibly affecting the caching efficiency and thus the optimal number of cells in a subgrid.

\paragraph{Forced scattering} The optical depths in dust simulations are often quite low ($\lesssim 1$), causing many photon packets to escape the spatial domain without interacting with the medium. The \emph{forced scattering} technique \citep{1959_Cashwell} overcomes this problem by forcing an interaction within the simulation box. The random interaction optical depth for a photon packet is sampled from a truncated exponential distribution and the photon packet weight is adjusted to compensate. This requires knowledge of the total optical depth along the photon packet's path, while this knowledge is only available after propagating the photon packet. To avoid performing the grid traversal twice, codes using this technique usually record details for the complete path so that this information can be used to determine both the total optical depth and the actual scattering location. Within our distributed-grid algorithm, however, recording information about each cell crossed by a photon packet is impractical because it would massively increase the memory footprint of the photon packet buffers. It would be straightforward and possibly even more efficient to perform the grid traversal twice. One can store the photon packets leaving the simulation box at the end of the preliminary propagation step in dedicated buffers and create a new task to relaunch these photon packets for the actual propagation step.

\paragraph{Peel-off} Other than determining the radiation field in the simulation box, one is often interested in creating synthetic observations of the simulated model, such as spectra or images, from a given line of sight.  To create these observables, the \emph{peel-off} technique \citep{1984_YusefZadeh} is used, whereby a photon packet with an appropriate weight is sent towards the observer during each regular emission and scattering event. In an extended version of our task-based algorithm, peel-off photon packets could be handled by the propagation task that also handles the preliminary photon packets in the forced scattering technique. When a peel-off photon packet reaches the boundary of the simulation box, it is then handed off to a task that will record its contribution to the observation. Because multiple such tasks may run at the same time, the recording data structure must be accessed in a thread-safe way. This can be accomplished through some form of locking, perhaps using atomic operations, or by providing a private copy of the data structure for each execution thread.

\paragraph{Thermal emission} Once the radiation field in each cell has been determined during a first photon packet emission phase, the corresponding dust emission spectrum can be calculated. The spectrum can be obtained either from the energy balance equation, if assuming local thermal equilibrium, or by determining the temperature probability distribution of a set of representative dust grains \citep{2015_Camps}. Because the calculation depends solely on information stored for each cell (in addition to constant dust material properties), it can be performed for each subgrid separately and thus easily in parallel. The same task can also act as a thermal emission source, generating new photon packets from the spectrum for each cell and placing them into a buffer associated with the local subgrid.

\subsection{Resonant line transfer}

The basic photon packet life cycle for tracing a resonant emission line such as Lyman-$\alpha$ through an interstellar medium is very similar to the cycle described for our algorithm in \sectionref{sec:algorithm}. Usually the relevant gas densities, chemical abundances and ionization states are considered to be known in advance of the simulation. As a result, there is no need to track the local radiation field during the simulation, simplifying that part of the algorithm. Often the main interest is to calculate the overall attenuation and the precise line shape for a given line of sight. This is usually accomplished using the peel-off technique, which can be implemented as a straightforward extension of the task-based algorithm as described in \sectionref{sec:dustRT}.

There is, however, a complication related to the high optical depths encountered by photon packets with a wavelength near the line resonance. Such a photon packet can easily experience many thousands of scatterings in a very compact region before its wavelength shifts sufficiently far from the line centre to allow escape. If this `scattering region' happens to straddle a subgrid boundary, the photon packet will be sent back and forth between subgrids countless times, significantly degrading the performance of the algorithm. This issue can be overcome by making subgrids \emph{overlap} each other at their respective boundaries, perhaps by the size of a cell. As long as the state of the medium is constant during the simulation (i.e.\ no radiation field tracking), this is fairly trivial to implement. Otherwise, the information recorded for the overlapping cells needs to be aggregated at the end of each photon packet propagation phase.

\subsection{Radiation hydrodynamics}
\label{sec:RHD}

\cmacionize{} can also be used as a full radiation hydrodynamics code. In this configuration, the grid structure used for the radiation phase is also used by the finite volume solver in the hydrodynamics phase. Our task-based implementation is very similar to that in \textsc{swift} \citep{2016Schaller}, i.e.\ appropriate hydrodynamical tasks are created for each subgrid and for the interactions between neighbouring subgrids. The radiation and integration phases are separated to avoid any issues caused by the differences in scheduling strategies. 

As in \cmacionizeone{}, the radiation phase can be scheduled after every time step or at regular simulation time intervals. The coupling between the two phases is affected by a single loop over all cells that updates the pressure or the energy according to the neutral fractions in the cell. Moreover, \cmacionize{} also supports a more elaborate coupling scheme, whereby the heating caused by photoionization is directly used as an energy source term, and is balanced by a temperature-dependent cooling rate given by interpolation
on pre-computed cooling tables. These various coupling modes can be selected at
run time by adjusting the parameter file.

\subsection{Spatial grids}

The algorithm presented in \sectionref{sec:algorithm} assumes a cuboidal spatial domain partitioned into subgrids according to a regular Cartesian grid, with subgrids that in turn have cells on a regular grid. Load imbalances between spatial regions can be addressed by duplicating the relevant subgrids (see \sectionref{subgrid_copies}). However, there are no provisions for varying the grid resolution depending on the density of the medium or on other local characteristics of the simulated model.

An obvious solution is to place the subgrids on the leaf nodes of an octree or some other hierarchical grid similar to the adaptive mesh refinement (AMR) grids used by many hydrodynamical codes. This requires adjustments to the methods for selecting the appropriate neighbour when a photon packet leaves a subgrid, and for locating the subgrid containing the starting position of a photon packet to begin with. Other than this, however, the task-based algorithm would remain largely the same.

One could also allow the individual subgrids to be subdivided into cells using a different type of grid, for example an octree or even an unstructured Voronoi mesh. However, the extra computational complexity and the larger subgrid memory footprint caused by such schemes would most likely eradicate the efficiency benefits of the presented task-based algorithm. 

In the current implementation of \cmacionize{}, the task-based algorithm is restricted to the regular grids described in \sectionref{sec:algorithm}. However, it is still possible to run \cmacionize{} using the non task-based version of the algorithm previously provided by \cmacionizeone{}, which supports both AMR and Voronoi grids.

\subsection{Distributed memory}
\label{sec:distributed_memory}

The algorithm described in \sectionref{sec:algorithm} assumes that all parallel execution threads can access a common memory space. This implies that the algorithm runs in a shared memory environment on a single compute node and is thus limited to the memory available on a single node. For a typical state-of-the-art node this means a memory size of $\sim{}$200~GB or a grid resolution of approximately $512^3$ cells. Larger memory sizes (up to $\sim{}$10~TB) are available, and we have successfully tested the algorithm using a $1024^3$ grid on a 1.5~TB node.

To support larger grids, however, the task-based algorithm would need to be ported to a distributed memory environment, where each execution thread (or small set of threads) runs in a separate \emph{process} with its own private memory space. These processes can be distributed over multiple compute nodes and they cooperate by exchanging messages over a network interconnect. While this setup allows combining the resources of a potentially large number of compute nodes, it requires an explicit decomposition of the data structures and the computational work involved in the algorithm, combined with a suitable communication strategy to exchange information and synchronise the different processes.

In such an environment, subgrids (or subgrid copies) would be distributed among the available processes. Radiation sources or copies of sources can likewise be distributed. Neighbouring subgrids may now reside in a different process, so that photon packets crossing the corresponding subgrid boundary need to be sent to the corresponding process before a subsequent photon packet propagation task can be created. At least in principle, this does not require many changes to the heart of our task-based MCRT algorithm, because the infrastructure for partitioning the spatial domain and for storing  and exchanging photon
packets is already in place. The adjustments are essentially of a technical nature. For example, one needs a new task type responsible for sending photon packet buffers to another process, and a mechanism that regularly checks for incoming messages as part of the routine that obtains a new task for a
thread.

The most important challenge, though, is to devise a strategy for distributing subgrids and radiation sources among the processes and compute nodes in such a way that the total computational load per compute node is balanced. We have experimented with a distributed memory version of the task-based
algorithm and verified that it indeed works. However, implementing an appropriate load-balancing strategy is outside the scope of this work, and \cmacionize{} does not include a distributed memory
version of the task-based algorithm.

\section{Conclusion}
\label{sec:conclusion}

Modern hardware architectures have a complex memory layout that favours algorithms with predictable and localised memory access patterns. This is especially true for MCRT simulations, which require a limited amount of calculation relative to the size of the data structures being accessed, and hence are strongly memory-bound. Using a novel implementation in the photoionization code \cmacionize{}, we showed that the MCRT algorithm can be adjusted to produce a much more optimal memory access pattern by subdividing the grid used to discretize the physical domain into many smaller subgrids. The standard photon packet propagation phase of the MCRT algorithm can be reconstructed by executing separate propagation tasks for these subgrids and storing photon packets that transition between subgrids in temporary buffers. The resulting algorithm has the same accuracy as a traditional algorithm, but is 2 to 4 times faster. The new algorithm shows good strong scaling on shared-memory systems, and has been successfully used on systems with up to 64 shared-memory threads, and with grid sizes of up to $1024^3$. We showed that optimal performance is the result of a trade-off between improved cache efficiency and increased overhead, mainly determined by the size and number of subgrids. We derived a heuristic model to determine the expected memory requirements for a simulation.

The new algorithm can be extensively analysed through run time diagnostics that are available without noticeable overhead, which allows exposing bottlenecks and load imbalances. For example, subgrids that experience an above-average computational load can be duplicated to increase concurrency and restore proper load balancing. This illustrates how the new algorithm separates physical from technical aspects, and allows addressing the latter without interfering with the former.

We also discussed how our new algorithm can be adapted for other MCRT applications, e.g.\ dust radiative transfer and resonant line transfer. We found that many existing techniques can be ported to our task-based approach with small modifications, so that the algorithm presented here is generally applicable. More fundamentally, perhaps, we showed that optimising memory access patterns in a memory-bound algorithm can yield significant performance gains.

\section*{Acknowledgements}

BV acknowledges partial support by STFC grant ST/M001296/1, and currently receives funding from the Belgian Science Policy Office (BELSPO) through the PRODEX project ``SPICA-SKIRT: A far-infrared photometry and polarimetry simulation toolbox in preparation of the SPICA mission'' (C4000128500). During development use was made of the DiRAC Data Intensive service at Leicester, operated by the University of Leicester IT Services, which forms part of the STFC DiRAC HPC Facility (www.dirac.ac.uk). The equipment was funded by BEIS capital funding via STFC capital grants ST/K000373/1 and ST/R002363/1 and STFC DiRAC Operations grant ST/R001014/1. DiRAC is part of the National e-Infrastructure. This work also used the DiRAC@Durham facility managed by the Institute for Computational Cosmology on behalf of the STFC DiRAC HPC Facility (www.dirac.ac.uk). The equipment was funded by BEIS capital funding via STFC capital grants ST/P002293/1, ST/R002371/1 and ST/S002502/1, Durham University and STFC operations grant ST/R000832/1. DiRAC is part of the National e-Infrastructure. Additional computational resources (Stevin Supercomputer Infrastructure) and services used in this work were provided by the VSC (Flemish Supercomputer Center), funded by Ghent University, FWO and the Flemish Government -- department EWI.

\bibliographystyle{aa}
\bibliography{main}

\end{document}